\documentclass[prl,aps,twocolumn,amssymb,floatfix,nofootinbib,superscriptaddress,notitlepage]{revtex4-1}
  \usepackage{times}
\usepackage{graphicx} 

\usepackage{amsmath}
\usepackage{amstext}
\usepackage{amssymb}
\usepackage{xfrac}
\usepackage[colorlinks,citecolor=blue]{hyperref}
\usepackage{graphicx}
\usepackage{amsmath}
\usepackage{amstext}
\usepackage{amssymb}
\usepackage{amsfonts}
\usepackage{longtable,booktabs}
\usepackage{hyperref}\usepackage{url}

\usepackage{amsbsy}
\usepackage{dcolumn}
\usepackage{amsthm}
\usepackage{bm}
\usepackage{multirow}
\usepackage{hyperref}
\hypersetup{
    colorlinks=true,
    linkcolor=magenta,
    filecolor=magenta,
    urlcolor=cyan,
}

\usepackage{mathrsfs}
\usepackage{amsfonts}
\usepackage{amsbsy}
\usepackage{dcolumn}
\usepackage{bm}
\usepackage{multirow}
\usepackage{color}

\usepackage{booktabs}
\usepackage{graphicx}
\usepackage{amsmath}
\usepackage{amstext}
\usepackage{amssymb}
\usepackage[colorlinks,citecolor=blue]{hyperref}
\usepackage{graphicx}
\usepackage{amsmath}
\usepackage{amstext}
\usepackage{amssymb}
\usepackage{amsfonts}
\usepackage{longtable,booktabs}
\usepackage{hyperref}\usepackage{url}
\usepackage{subfigure}%
\usepackage{dsfont}
\usepackage{datetime}

\usepackage{amsbsy}
\usepackage{dcolumn}
\usepackage{amsthm}
\usepackage{bm}
\usepackage{esint}
\usepackage{multirow}
\usepackage{hyperref}
\hypersetup{
    colorlinks=true,
    linkcolor=magenta,
    filecolor=cyan,
    urlcolor=blue,
}
\usepackage{booktabs}

\usepackage{ctable}
\usepackage{cleveref}
\usepackage{comment}
\usepackage{mathrsfs}
\usepackage{amsfonts}
\usepackage{amsbsy}
\usepackage{dcolumn}
\usepackage{bm}
\usepackage{multirow}
\usepackage{color}
\newcounter{defcounter}
\usepackage{amsthm}

\def\Z{\mathbb{Z}}

  \usepackage{extarrows}
 \usepackage{graphicx}
    \usepackage{amssymb}
    \usepackage{amsthm}
    \usepackage{mathtools}
    \usepackage{adjustbox}

\def\MM{{\cal M}}

\def\PP{\partial}
\def\SS{{\cal S}}
\usepackage{hyperref}
\hypersetup{
    colorlinks=true,
    linkcolor=blue,
    filecolor=blue,
    urlcolor=blue,
}
\def\Z{\mathbb{Z}}

\begin{document}
\title{Braiding with Borromean Rings in (3+1)-Dimensional Spacetime}

\author{AtMa P.O. Chan}
\affiliation{Department of Physics and Institute for Condensed Matter Theory, University of Illinois at Urbana-Champaign, Illinois 61801, USA}

\author{Peng Ye}
\email{yepeng5@mail.sysu.edu.cn}
\affiliation{School of Physics, Sun Yat-Sen University, Guangzhou, 510275, China}
\affiliation{Department of Physics and Institute for Condensed Matter Theory, University of Illinois at Urbana-Champaign, Illinois 61801, USA}

\author{Shinsei Ryu}
\email{ryuu@uchicago.edu}
\affiliation{James Franck Institute and Kadanoff Center for Theoretical Physics,
University of Chicago, Illinois 60637, USA}

\begin{abstract}
While winding a particle-like excitation around a loop-like excitation yields the celebrated Aharonov-Bohm phase, we find a distinctive braiding phase in the absence of such mutual winding. In this work, we propose an exotic particle-loop-loop braiding process, dubbed the \emph{Borromean-Rings braiding}. In the process, a particle moves around two unlinked loops, such that its trajectory and the two loops form the Borromean-Rings or more general Brunnian links. As the particle trajectory does not wind with any of the loops, the resulting braiding phase is fundamentally different from the Aharonov-Bohm phase. We derive an explicit expression for the braiding phase in terms of  the underlying Milnor's triple linking number. We also propose Topological Quantum Field Theories consisting of an $AAB$-type topological term which realize the braiding statistics.
\end{abstract}

\maketitle

\textit{Introduction}---Braiding statistics is a quantum mechanical phenomenon in which a quantum state acquires a holonomy when winding an excitation adiabatically around other excitations \cite{Leinaas1977,PhysRevLett.48.1144,PhysRevLett.49.957}. It  arises from the ambiguous weightings for distinct homotopy classes of trajectories in the Feynman's path integral which sums over all continuous paths in the configuration space \cite{wilczek1990fractional,wu84}.
Not only is quantum statistics an important subject in fundamental physics,
it is also a crucial data in characterizing topological order in long-range entangled phases of matter \cite{Wen1995,wen2004quantum,string1.5}. Moreover, braiding statistics has been recently shown to be a powerful diagnostic of Symmetry-Protected Topological (SPT) phases \cite{levin_gu_12, Chenlong,PhysRevB.85.075125,1DSPT,Chen_science}. By now, braiding statistics in (2+1)D has been thoroughly studied through the braid group and formulated in the theory of anyons \cite{wen_stacking,Kitaev2006,wang,BONDERSON20082709}. Nevertheless, our understanding of braiding statistics in (3+1)D is still far from mature. The core reason is that the possible loop excitations complicate the
configuration space in the path integral. While the simplest particle-particle braiding is always trivial due to the contractibility of particle trajectories around the other particle, the possible braiding statistics is significantly enriched if the spatially extended loop excitations are taken into account. The most well-known example is the particle-loop braiding statistics in which a particle carried along a non-contractible cycle around a loop experiences the Aharonov-Bohm effect \cite{abeffect}.

\begin{figure}[t]
\centering
\includegraphics[scale=0.28]{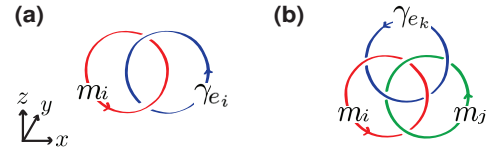}
\caption{(Color online) (a) Particle-loop braiding: a particle $e_i$ travels around a loop $m_i$ such that the braiding trajectory $\!\gamma_{e_i}\!$ and $m_i$ form a Hopf link. (b) Borromean-Rings braiding: a particle $e_k$ moves around two unlinked loops $m_i, m_j$ such that $m_i$, $m_j$ and the trajectory $\!\gamma_{e_k}\!$ form the Borromean rings (or generally the Brunnian link).
}\label{fig1}
\end{figure}

The peculiar braiding phase in the Aharonov-Bohm effect has been understood to be associated with the winding between the particle trajectory and the loop \cite{Leinaas1977,PhysRevLett.48.1144,PhysRevLett.49.957} [Fig.~\ref{fig1}(a)]. In this work, we argue that the statistical interaction between particles and loops can appear in a more general context. We consider the effect of braiding a particle with more than one loop. Importantly, we find that there can be a non-trivial braiding phase even without winding the particle around any loop [Fig.~\ref{fig1}(b)].\! Particularly, in braiding a particle around two unlinked loops, a braiding phase appears when the particle trajectory and the loops form a Brunnian link, which is formed by three mutually unlinked circles. For example, the simplest Brunnian link is the Borromean-Rings link. While the traditional particle-loop braiding statistics is dictated by the Hopf linking number $\mathfrak{L}$, we show that the \emph{particle-loop-loop braiding} statistics is instead governed by a higher order linking number, called the Milnor's triple linking number $\bar{\mu}$ \cite{milnor1954link}.

Physically, braiding statistics involving particles and loops can be realized in Abelian discrete gauge theories. For example, non-trivial particle-loop braiding statistics can be realized in $\Z_N$ gauge theory, which describes the deconfined phase of (3+1)D type-II superconductor with a charge-$N$ condensate \cite{fradkin2013field,hansson2004superconductors}. In such a gapped phase of matter, the excitation spectrum is generated by a particle $e$ and a loop $m$ under fusion, where $Ne, Nm$ are both trivial. Carrying a particle $e$ in a closed path $\gamma_e$ around a loop $m$ leads to the quantized phase $\frac{2\pi}{N}\mathfrak{L}(m,\gamma_e)$ [Fig.~\ref{fig1}(a)]. Recently, more exotic multi-loop braiding statistics and particle statistical transmutation have been demonstrated in discrete gauge theories with larger gauge group $G\!=\!\prod_i \Z_{N_i}$ \cite{wang_levin1,jian_qi_14,string5,PhysRevX.6.021015,string6,YeGu2015,corbodism3,ye16_set,2016arXiv161008645Y,string4,PhysRevLett.114.031601,3loop_ryu,string10,2016arXiv161209298P,Tiwari:2016aa}, which is a system of $\Z_{N_i}$ gauge theories with a collection of flavors $i\!\in\!\mathcal{F}$. In these theories, the vacuum expectation of any physical braiding process yields a complex phase factor.

\begin{figure}[t]
\centering
\includegraphics[scale=0.28]{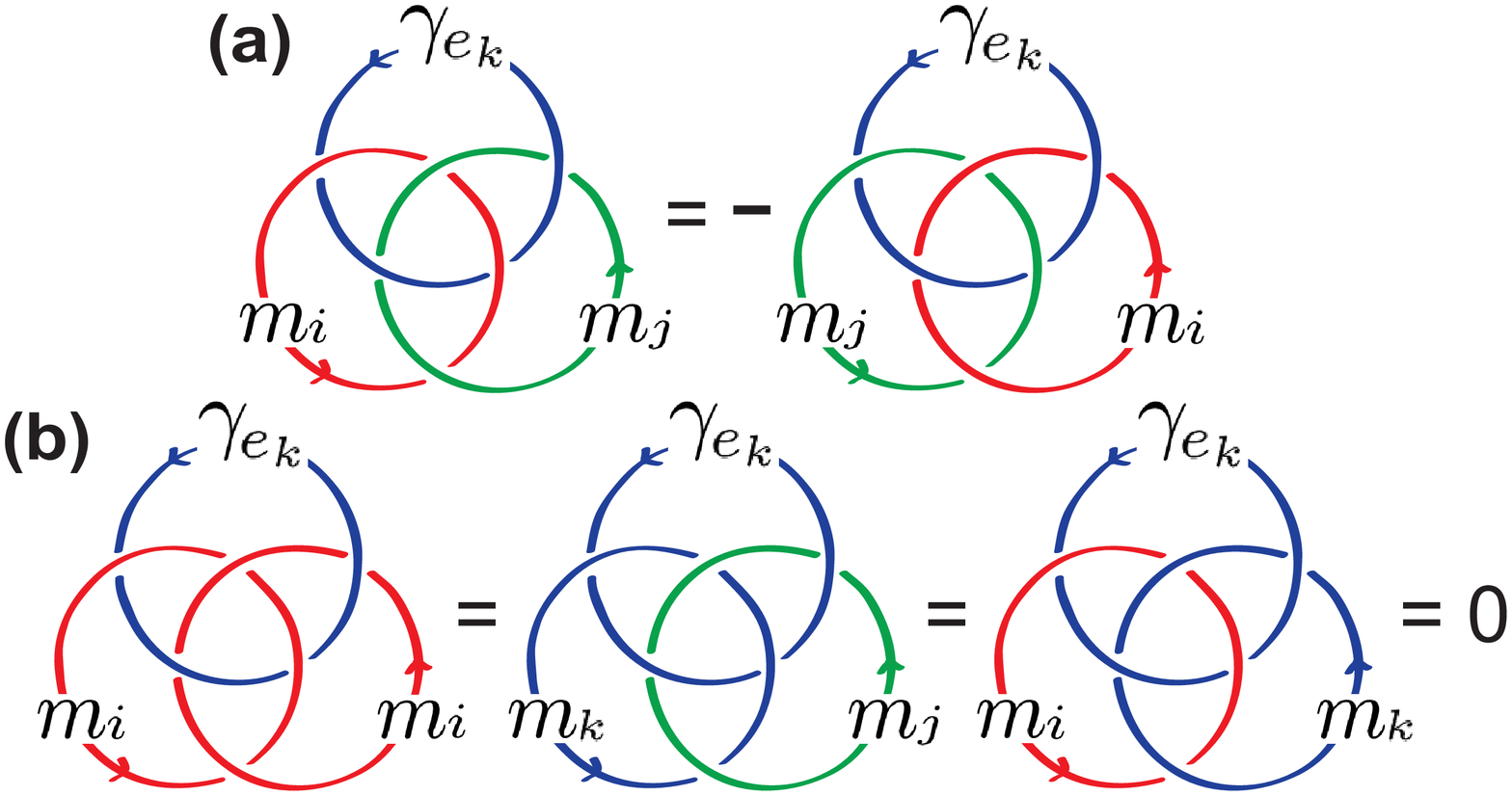}
\caption{(Color online) Constraints on the BR braiding phase $\mathsf\Theta(L)$. (a) $\mathsf{\Theta}(L)$ changes sign if $m_i$ and $m_j$ are exchanged. (b) $\mathsf{\Theta}(L)$ vanishes if any two of three components  belong to the same gauge group.}\label{fig2}
\end{figure}

In this work, we introduce the Borromean-Rings (BR) braiding, namely, the particle-loop-loop braiding generating the Brunnian links [Fig.~\ref{fig1}(b)], in discrete gauge theories with $G\!=\!\prod_i \Z_{N_i}$. Contrary to the particle-loop braiding where the particle trajectory is linked with the loop, the particle trajectory is not linked to any of the two loops in the BR braiding. By following a line of geometric arguments, we derive constraints [Fig.~\ref{fig2}] and quantization condition of the BR braiding phase if it exists. Then we obtain an explicit formula for the braiding phase, which is expressed in terms of the Milnor's triple linking number [Eq.~(\ref{BR_geometric})]. Also, we construct Topological Quantum Field Theories (TQFTs) with a $BF$ action dressed with an $AAB$ topological term ($A$ and $B$ denote some 1-form and 2-form gauge fields respectively) [Eq.\ (\ref{TQFT})] which support non-trivial BR braiding statistics [Eq.\ (\ref{br_eff_ective})]. The resulting BR braiding phase agrees with the result from geometric arguments. This work is concluded with several remarks and future directions.

\textit{Preliminaries}---As a warm-up, we discuss general aspects of braiding a particle around loops in Abelian discrete gauge theories. Here, we are primarily interested in the classes of closed paths which could lead to non-trivial braiding statistics \cite{wilczek1990fractional,wu84}. As the particle travels in the complement of loops, the braiding statistics must be trivial if its closed trajectory can be adiabatically shrunk to a point. Equivalently, if the trajectory and the loops are viewed as a link, the braiding statistics is trivial if the trajectory can be unlinked from the loops. Under the deformation, the trajectory can cross with itself since the intersection point corresponds to the particle position at different time instances which can never interact. However, it cannot cross with any of the loops since the Aharonov-Bohm effect can contribute to braiding statistics. Besides, the loops can also undergo adiabatic deformation. Note that while Aharonov-Bohm interaction is possible among loops, there is no Aharonov-Bohm self-interaction \cite{PRESKILL199050,PhysRevLett.64.1632,ALFORD1992251,BUCHER19923}. In other words, each loop is allowed to cross with itself but not with other loops. Under such \emph{link homotopy}, each link component can cross with itself but not with other link components \cite{milnor1954link}. Any particle trajectory that cannot be shrunk to a point under link homotopy can in principle lead to a non-trivial braiding phase.

In such formulation, each homotopy class of links is assigned with a braiding phase that depends only on the underlying linking numbers. Hence, while the particle-loop braiding phase is determined by the Hopf linking number $\mathfrak{L}$, the particle-loop-loop braiding phase is governed by the three mutual Hopf linking numbers and the Milnor's triple linking number $\bar{\mu}$ \cite{milnor1954link}. In this work, we study the particle-loop-loop braiding statistics which cannot be simply explained by the Hopf linking numbers. Nevertheless, the higher order linking number $\bar{\mu}$ is an invariant under link homotopy iff all the three mutual Hopf linking numbers vanish. So when the trajectory and the two loops are mutually unlinked, we expect a well-defined braiding phase determined by $\bar{\mu}$.

\textit{Borromean-Rings Braiding}---Here we introduce the BR braiding in the context of Abelian discrete gauge theories. Pick any $i,j,k\in \mathcal{F}$, the BR braiding is a particle-loop-loop braiding in which a $\Z_{N_k}$ particle $e_k$ is carried around mutually unlinked $\Z_{N_i}$ loop $m_{i}$ and $\Z_{N_j}$ loop $m_j$ such that the closed path and the two loops form the Borromean-Rings, or generally the Brunnian link [Fig.~\ref{fig1}(b)]. Let $\mathsf{B}_{i,k}$ and $\mathsf{B}_{j,k}$ be the quantum operators of braiding $e_k$ around $m_{i}$ and $m_j$ respectively. Given the two loops, any braiding process can be written as a sequential operation in $\mathsf{B}_{i,k}$ and $\mathsf{B}_{j,k}$ as well as their inverses, in which $m_{i}$ and $m_j$ together with the braiding trajectory $\gamma_{e_{k}}$ can be viewed homotopically as a link $L$. For the BR braiding, since the braiding trajectory is not linked with any of the two loops, the sum of exponents is zero for both $\mathsf{B}_{i,k}$ and $\mathsf{B}_{j,k}$. For example, the braiding process giving Borromean-Rings link is written as $\mathsf{B}_{j,k}^{-1}\mathsf{B}_{i,k}^{-1}\mathsf{B}^{}_{j,k}\mathsf{B}^{}_{i,k}$ \cite{AlTopo}. Since the exponent sum of each of them is zero, if any of the two constituent braidings $\mathsf{B}_{i,k}$ and $\mathsf{B}_{j,k}$ gives only an Abelian phase, the BR braiding statistics must be trivial. Hence, non-trivial BR braiding statistics implies that $m_{i}, m_j$ and $e_k$ support non-Abelian braiding statistics, despite the Abelianess of the gauge group $G$. We denote the overall BR braiding phase as $\mathsf{\Theta}(L)$. Below, we are going to extract several constraints on the BR braiding phase geometrically.

First of all, since the BR braiding generates the Brunnian links, the geometric properties of the braiding phase are dictated by the Milnor's triple linking number $\bar{\mu}$. Let $c_{ij,k}$ be the braiding phase for the simplest BR braiding of particle $e_k$ around $m_{i}$ and $m_j$ forming Borromean-Rings with $\bar{\mu}\!=\!1$. Consider carrying the particle $e_{k}$ along its original path $\!\gamma_{e_{k}}\!$ by $w$ times. It generates a Brunnian link $L$ with $\bar{\mu}\!=\!w$. Generally, any Brunnian link with $\bar{\mu}\!=\!w$ can be generated this way up to link homotopy. Since the BR braiding is operated repeatedly, the braiding phase accumulates over $w$ times. Therefore the BR braiding phase should take the linear form $\mathsf{\Theta}(L)\!=\!c_{ij,k}\bar{\mu}({m_{i}}, {m_{j}},\gamma_{e_{k}})$, where the three entries for $\bar{\mu}$ are respectively the first, second and third component of $L$. Physically, $c_{ij,k}$ encodes the braiding data while $\bar{\mu}$ gives the geometric properties of the BR braiding process.

Next, we demonstrate the anti-symmetry of the BR braiding statistics. Consider braiding a particle $e_k$ around $m_i$ and $m_j$ in a way generating the Borromean-Rings [Fig.~\ref{fig2}(a)]. Viewing the process from the opposite side, the particle $e_k$ travels around loops $m_{j}$ and $m_{i}$ with the orientation of the three components are flipped. Since reversing the orientation of any component of $L$ causes $\bar{\mu}$ to change its sign, flipping the orientation of the three components gives a minus sign to the braiding phase. Hence braiding around $m_{i}$ and $m_{j}$ is minus the braiding around $m_{j}$
and $m_{i}$, so $c_{ji,k}\!=\!-c_{ij,k}$. Such sign change can also be understood as coming from braiding $e_k$ around $m_{i}$ and $m_{j}$ but with flipped labels $m_i$ and $m_j$ in $L$, which changes the sign of $\bar{\mu}$. Generally, for arbitrary Brunnian link $L$, if
the labels $m_i$ and $m_j$ are interchanged,
$\mathsf\Theta(L)\!\rightarrow\! c_{ij,k}\bar{\mu}({m_{j}},{m_{i}},\gamma_{e_{k}})\!=\!-\mathsf{\Theta}(L)$.
Thus interchanging the labels of $m_i$ and $m_j$ flips the sign of $\mathsf{\Theta}(L)$.

Next, we show that the BR braiding phase vanishes if any two objects involved are from the same gauge group. Consider the decomposition $\mathsf{B}_{j,k}^{-1}\mathsf{B}_{i,k}^{-1}\mathsf{B}^{}_{j,k}\mathsf{B}^{}_{i,k}\!=\!e^{ic_{ij,k}}$ for the BR braiding generating the Borromean-Rings link [Fig.~\ref{fig2}(b)]. If $i\!=\!j$, the product of operators reduces to identity and hence $c_{ii,k}\!=\!0$. If $i\!=\!k$, since $\mathsf{B}_{i,k}$ is guaranteed to give an Aharonov-Bohm phase $\frac{2\pi}{N_k}$ under the $\Z_{N_k}$ gauge group, the product of operators again reduces to identity and hence $c_{kj,k}\!=\!0$. Similarly, we also have $c_{ik,k}\!=\!0$. Consequently,  $\mathsf{\Theta}(L)$ vanishes if any two of the indices in $L$ are identical. In other words, non-trivial BR braiding appears only for distinct indices $i,j,k$. In particular, non-trivial BR braiding implies non-Abelian particle-loop braidings $\mathsf{B}_{i,k}, \mathsf{B}_{j,k}$ for \emph{distinct} flavors, rendering each of them to be gauge non-invariant.

We now derive the quantization rule of the BR braiding phase. Consider the BR braiding forming the Borromean-Rings with braiding phase $c_{ij,k}$. Imagine scaling up the phase by $N_k$ by carrying $e_{k}$ along $\gamma_{e_k}$ repeatedly for $N_k$ times. The whole process is equivalent to carrying $N_k e_{k}$ once along $\gamma_{e_k}$. Since $N_k e_{k}$ is a trivial particle, and braiding is compatible with fusion, we have $N_{k}c_{ij,k}\!=\!0\textrm{ mod }2\pi$. Now imagine scaling up the phase by $N_i$ by winding $m_i$ along its locus for $N_i$ times instead. Again, since $N_i m_i$ is trivial and braiding is compatible with fusion, we have $N_{i}c_{ij,k}\!=\!0\textrm{ mod }2\pi$. Similarly, $N_{j}c_{ij,k}\!=\!0\textrm{ mod }2\pi$. Combining all the three conditions, we have
$c_{ij,k}\!=\!\frac{2\pi k_{ij,k}}{N_{ijk}}$, where $k_{ij,k}$ is an integer and $N_{ijk}$ denotes
the greatest common divisor of $N_i, N_j$ and $N_k$. Finally, we get the formula for the BR braiding phase
\begin{align}
\mathsf{\Theta}(L)=\frac{2\pi k_{ij,k}}{N_{ijk}}\bar{\mu}({m_{i}},{m_{j}},\gamma_{e_{k}})\,,\label{BR_geometric}
\end{align}
where all properties of the coefficient $c_{ij,k}$ propagate to $k_{ij,k}$. That is, $k_{ji,k}\!=\!-k_{ij,k}$ and $k_{ij,k}$ vanishes if any of the two indices are the same. Since $\mathsf{\Theta}(L)$ is defined up to $2\pi$, the parameter $k_{ij,k}\!\in\!\Z_{N_{ijk}}$. We conclude one of our main results: if the BR braiding statistics exists in discrete gauge theories, the braiding phase must take the form as Eq.\ (\ref{BR_geometric}).
Below, we construct explicitly field-theoretic models which support non-trivial BR braiding statistics.

\textit{TQFTs with $AAB$ Topological Term}--- It is believed that
low energy physics of long-range entangled phases of matter is captured by some
TQFTs \cite{sarma_08_TQC}. For example, the topological features of discrete gauge theories with $G\!=\!\prod_i\Z_{N_i}$ are known to be described by the $BF$ theories with action
$S_{\rm BF}\!\!=\!\! \int\sum_i\frac{N_i}{2\pi} B^i  dA^i\!$, where the 1-form $A^i$ and 2-form $B^i$ are compact U(1) gauge fields describing the loop and particle degrees of freedom respectively \cite{horowitz89,Baez2011}. The $\!\Z_{N_i}\!$ fusion structure of particles and loops is encoded in the cyclic Wilson integrals of $A^{i}$ and $B^{i}$. Moreover, the Aharonov-Bohm effect is captured by the effective action $S_{\rm Hopf}\!\!=\!\!\sum_i\!\frac{2\pi}{N_{i}}\!I_{\rm Hopf}[\Sigma^i\!,J^i]$, where the 3-form $J^i$ and 2-form $\Sigma^i$ are respectively the particle and the loop sources describing the braiding process, and  $I_{\rm Hopf}[\Sigma^i\!,J^i]\!\!=\!\!\int\!\Sigma^i d^{-1}\!J^i$ counts the Hopf linking $\mathfrak{L}(m_i,\gamma_{e_i})$. On top of the $BF$ theories, exotic braiding statistics can be realized by introducing an extra topological term \cite{PhysRevLett.114.031601,jian_qi_14,wang_levin1,string4,string5,YeGu2015,corbodism3,ye16_set,2016arXiv161008645Y,3loop_ryu,string10,2016arXiv161209298P,Tiwari:2016aa,string6,PhysRevX.6.021015,bti2,Kapustin2014,string7,Keyserlingk13,Walker2012}. Below, we introduce the $AAB$ term and show that the resulting  theories support the BR braiding statistics [Eq.\ (\ref{br_eff_ective})].

We begin by exhausting the possible $AAB$ terms for physical theories. Consider adding an
$\!A^iA^jB^k\!$ term with some real coefficient $\!c_{ij,k}\!$
upon the $BF$ theories [Eq.\ (\ref{TQFT})]. We are going to show that $\!c_{ij,k}\!$ here
satisfies the same set of constraints as that in the previous discussion. First,
notice that not all possible terms are independent, more specifically,
interchanging $A^i$ and $A^j$ gives the same term but with a minus sign, hence
$c_{ji,k}\!=\!-c_{ij,k}$. Second, some choices of indices are improper. We see $i\!\neq\!
j$, otherwise $\!A^iA^jB^k\!$ vanishes. For any flavor $i$, since either
$A^{i}$ or $B{^{i}}$ is reserved as the Lagrange multiplier which enforces the
$\Z_{N_i}$ fusion structure, $A^{i}$ and $B{^{i}}$ of the same flavor cannot
simultaneously appear on top of the $BF$ theories, so
$i,j\!\neq \!k$. In other words, $c_{ij,k}\!=\!0$ if any two indices are the same. So $G$ requires at least three $\Z_{N_i}$ group components for a legitimate $A^iA^jB^k$ term. Lastly, we show that $\!c_{ij,k}\!$ is quantized due to large gauge invariance. To this end, we pick some distinct $i,j,k\!\in\!\mathcal{F}$ for the $\!A^iA^jB^k\!$ term and focus on the three cyclic group components involved. Consider
\begin{align}\label{TQFT}
S=S_{\rm BF}+S_{\rm AAB}\,,\quad S_{\rm AAB}=\int\frac{n c_{ij,k} }{(2\pi)^3} A^i A^j B^k\,,\,
\end{align}
where $n=N_iN_jN_k$. Let $a, b=i,j$, the action $S$ is invariant up to a surface term under the gauge transformation
\begin{align}\label{gtrans}
\begin{split}
\mbox{$A^a\rightarrow A^a+d\alpha^a\,,\, B^a\rightarrow  B^a+d\beta^a+\mathcal{X}^a$}\,,\\
\mbox{$B^k\rightarrow B^k+d\beta^k\,,\, A^k\rightarrow A^k+d\alpha^k+\mathcal{X}^k$}\,,
\end{split}
\end{align}
where, to compensate the gauge change of the $S_{\rm AAB}$ term, the Lagrange multipliers $B^a$ and $A^k$ transform with extra twists
\begin{align}
\begin{split}
&\mbox{$\mathcal{X}^{a}=-\sum_{b}\frac{n c_{ab,k}}{(2\pi)^2 N_a}(\alpha^{b}B^k-A^{b}\beta^k+\alpha^{b}d\beta^k)$}\,,\\
&\mbox{$\mathcal{X}^k=-\sum_{ab}\frac{n c_{ab,k}}{(2\pi)^2N_k}(\alpha^a A^{b}+\frac{1}{2}\alpha^a d\alpha^b)$}\,.
\end{split}
\end{align}
After integrating out the Lagrange multipliers, the action $S$ reduces to $S_{\rm AAB}$, where $A^{a}$ and $B^k$ are enforced to be closed with cyclic Wilson integrals $\oint\! A^a\!\in\!\frac{2\pi}{N_a}\Z_{N_a}\!$ and $\oint\!B^k\!\in\!\frac{2\pi}{N_k}\Z_{N_k}\!$ over any closed manifolds. Under large gauge transformation, the gauge change of the action $S_{\rm AAB}$ consists of terms which take values in integral multiple of $\!N_{i}c_{ij,k},\!N_{j}c_{ij,k}\!$ and $\!N_{k}c_{ij,k}\!$ (SM Part 2.1.3 \cite{SM_cite}). The large gauge invariance of the resulting $S_{\rm AAB}$ term, which implies that $\!N_{i}c_{ij,k},\!N_{j}c_{ij,k}\!$ and $\!N_{k}c_{ij,k}\!$ vanish mod $2\pi$, leads to the desired coefficient quantization $c_{ij,k}\!=\!\frac{2\pi k_{ij,k}}{N_{ijk}}\!$ for integral $k_{ij,k}$.

\begin{figure}[h]
\centering
\includegraphics[scale=0.28]{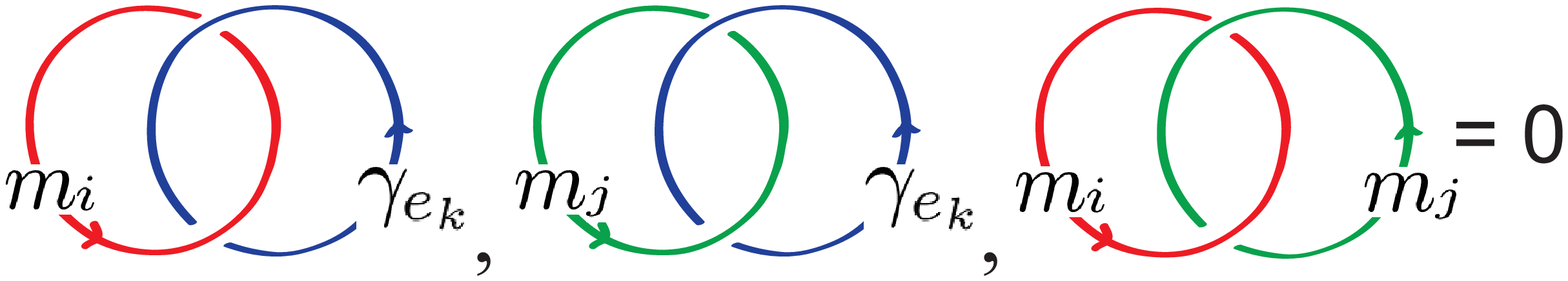}
\caption{(Color online) Illustration of the braiding constraints in Eq.~(\ref{two_constraints_11}). If the loops $m_{i}$ and $m_{j}$ are static, then $m_{i}$, $m_{j}$ and $\gamma_{e_k}$ are mutually unlinked circles for gauge invariant braiding process.}\label{fig3}
\end{figure}

Next, we discuss the constraints on the braiding. Consider a generic
braiding described by some closed world lines for particles and closed world sheets for loops.
The corresponding conserved particle sources $J^{i}, J^{j}$ and $J^{k}$, and loop sources $\Sigma^{i}, \Sigma^{j}$ and $\Sigma^{k}$ can be incorporated into $S$ via the source term
\begin{align}
S_{\rm s}
=-\!\int\!\textrm{\small$\sum_{a}$}\big(J^a\!A^a+\Sigma^a\mathcal{B}^a\big)+\Sigma^kB^k+J^k\mathcal{A}^k,
\end{align}
where $a\!=\!i,j$. The sources $\Sigma^a\!$ and $J^k\!$ are respectively coupled to the modified Lagrange multipliers $\mathcal{B}^a$ and $\mathcal{A}^k$ defined as
\begin{align}
\begin{split}
&\mbox{\,\,\,\,\,\,\,\,\,\,$\mathcal{B}^a=B^a-\sum_{b}\frac{nc_{ab,k}}{2(2\pi)^2 N_a}(A^bd^{-1}\!B^k\!-\!d^{-1}\!A^b\,B^k)$}\,,\\
&\mbox{\,\,\,\,\,\,\,\,\,\,$\mathcal{A}^k=A^k-\sum_{ab}\frac{nc_{ab,k}}{2(2\pi)^2 N_k}A^{a}d^{-1}\!A^b$}\,,
\end{split}
\end{align}
which transform like ordinary gauge fields. \!\!Under gauge transformation, $A^a$ and $B^k$ change by a pure gauge, so $J^a\!A^a$ and $\Sigma^k\!B^k$ must be gauge invariant. However, $\mathcal{B}^a$ and $\mathcal{A}^k$ change by a total derivative of non-local terms, which is not strictly a pure gauge, so $\Sigma^a\!\mathcal{B}^a$ and $J^k\!\mathcal{A}^k$ may not be gauge invariant for arbitrary braiding. Remarkably, $\!S_{\rm s}\!$ is gauge invariant iff
\begin{align}\label{two_constraints_11}
I_{\rm Hopf}[\Sigma^a,J^k]=0\,,\,\,\,\,\,\,
I_{\rm Hopf}[\Sigma^a,\hat{J}^b]=0\,\,\,\,(a\neq b)\,,
\end{align}
for any $\hat{J}^b$ describing current on the world sheet of $m_b$, for $a,b\!=\!i,j$
(SM Part 2.1.4 \cite{SM_cite}).
The first constraint means that the particle-loop braiding of $e_k$ and $m_a$ alone is not gauge invariant for $a=i,j$. The physical meaning of the second constraint can be understood by considering different choices of $\hat{J}^b$. Take $\hat{J}^b$ as the current of any point on $m_b$, it means that no point on the loop $m_b$ can braid around the loop $m_a$ for $a\neq b$. Take $\hat{J}^b$ as a time slice of the world sheet of $m_b$, which corresponds to the locus of $m_b$ at a fixed time, it means that there is no linking between the loops $m_a$ and $m_b$ for $a\neq b$. In particular, since crossing between two loops always changes their linking number, loop crossing of $m_a$ and $m_b$ is not gauge invariant  for $a\neq b$. If the loops are static, then the constraints in Eq.~(\ref{two_constraints_11}) simply mean that the loops $m_{i}$ and $m_{j}$ and the particle trajectory $\gamma_{e_k}$ must be mutually unlinked [Fig.~\ref{fig3}].

Now, we show that these theories support non-trivial BR braiding statistics. With the source term $S_{\rm s}$, the Lagrange multipliers $B^a$ and $A^k$ enforce that $\Sigma^{a}\!\!=\!\!\frac{N_a}{2\pi}dA^a$ and $J^{k}\!\!=\!\!\frac{N_k}{2\pi}dB^k$, for $a=i,j$. Consequently, $S+S_{\rm s}$ leads to the effective action
\begin{align}
S_{\rm eff}\!=\!S_{\rm Hopf}\!+\!S_{\rm BR}\,,\quad S_{\rm BR}\!\!=\!\!\frac{2\pi k_{ij,k}}{N_{ijk}}I_{\rm BR}[\Sigma^i\!,\Sigma^j\!,J^k]\,.
\end{align}
As in the $BF$ theories, $\!S_{\rm Hopf}\!$ accounts for the Aharonov-Bohm effect for particle-loop braiding within the same flavor. Here, the $AAB$ term induces an extra effect described by $\!S_{\rm BR}\!$ with
\begin{align}\label{IBF}
I_{\rm BR}[\Sigma^i\!,\Sigma^j\!,&J^k]
=\mbox{$\int  d^{-1}\Sigma^id^{-1}\Sigma^jd^{-1}J^k$}\nonumber\\
&\mbox{$- \frac{1}{2}\Sigma^i(d^{-1}\Sigma^jd^{-2}J^k-d^{-1}J^kd^{-2}\Sigma^j)$}\nonumber\\
&\mbox{$- \frac{1}{2}\Sigma^j(d^{-1}J^kd^{-2}\Sigma^i-d^{-1}\Sigma^id^{-2}J^k)$}\nonumber\\
&\mbox{$- \frac{1}{2}J^k(d^{-1}\Sigma^id^{-2}\Sigma^j-d^{-1}\Sigma^jd^{-2}\Sigma^i)$}\,.
\end{align}
Analogous to $I_{\rm Hopf}$ that counts the Hopf linking $\mathfrak{L}(m_i,\gamma_{e_i})$, the integral $I_{\rm BR}$ also admits a geometric interpretation (SM Part 2.2 \cite{SM_cite}).
Consider the gauge invariant particle-loop-loop braiding, where $e_k\!$ travels around two static loops $m_{i}\!$ and $m_{j}\!$ with mutually unlinked $m_{i},\!m_{j}\!$ and $\gamma_{e_k}$. $I_{\rm BR}$ counts the Milnor's triple linking number $\bar{\mu}({m_{i}},\!{m_{j}},\!\gamma_{e_{k}}\!)$. Hence,
\begin{align}
S_{\rm BR} = \frac{2\pi k_{ij,k}}{N_{ijk}}\bar{\mu}({m_{i}},{m_{j}},\gamma_{e_{k}})\,.\label{br_eff_ective}
\end{align}
In other words, the BR braiding process  produces a BR braiding phase $\mathsf{\Theta}(L)={S_{\rm BR}}$. The field-theoretic result here further justifies the main result (\ref{BR_geometric}) obtained independently by geometric arguments. By noting that the braiding phase $S_{\rm BR}$ is defined mod $2\pi$, we see $k_{ij,k}\!\in\!\Z_{N_{ijk}}$ can be used to classify discrete gauge theories with BR braiding statistics.

\textit{Conclusions}---In this work, we introduced the BR braiding statistics from both geometric arguments and  field-theoretic approach. Same quantum phenomenon is expected to appear also in discretized spacetime \cite{PhysRevB.95.155118,2017arXiv170200868B}. The BR braiding statistics reveals exotic phases with non-abelian topological order under Abelian gauge group $G$. Due to the duality correspondence between topological order and SPT order \cite{levin_gu_12,wang_levin1}, the proposed BR braiding statistics immediately implies a new class of SPT order with global symmetry $G$ (SM Part 3 \cite{SM_cite}).
In principle, the BR braiding phase can be observed by interferometry experiments similar to the measurements of Aharonov-Bohm effect, though the experimental design could be challenging. Nevertheless, it is expected to show up numerically as Berry phase in lattice Hamiltonian with higher form gauge symmetry \cite{PhysRevB.95.155118,2017arXiv170200868B}.
Lastly, it will be amusing to study entanglement properties of Eq.~(\ref{TQFT}) \cite{2017arXiv171011168W} and explore the fermionic analog of BR braiding statistics.


We are indebted to M.~Levin, X.~Chen, Y.-M.~Lu, C.~Wang, M.~Cheng, Y.-S.~Wu, R.~Thorngren, and E.~Fradkin for their thoughtful discussions. P.Y. thanks  G.~Chen for warm hospitality  in Fudan where part of the work was done. This work was supported in part by the National Science Foundation grant DMR~1455296 (A.P.O.C., S.R.) and DMR~1408713 (P.Y.) at the University of Illinois, grant of the Gordon and Betty Moore Foundation EPiQS Initiative through Grant No. GBMF4305 (P.Y.),  the startup grant in Sun Yat-sen University (P.Y.), and the Thousand Youth Talents Plan of China (P.Y.).
 
 \clearpage
 \onecolumngrid
\appendix

\begin{center}
\large\textbf{Supplemental Material}
\end{center}
 
 \twocolumngrid

\tableofcontents

\section{OUTLINE}

 \textbf{This supplemental material (SM) is devoted to work out the technical details for the main text of BR braiding statistics. While the overall main text is written in a self-contained manner, the SM here provides a deeper discussion to every little step. The SM is organized in a way parallel to the structure of the main text. Part 1 of the SM concerns the geometric picture of particle-loops braiding process. Part 2 of the SM elaborates the calculations in the TQFTs. In particular, we show the detailed proof for the coefficient quantization in the TQFTs in Part 2.1.3; we prove the constraints on the braiding process under gauge invariance in Part 2.1.4; we show the geometric meaning of the effective action in Part 2.2. Note that in order to better understand the geometric interpretation in Part 2.2, readers are recommended to go through the review for the geometric definitions of the linking numbers in Part 1.2. Finally, we add in Part 3 for the discussion of physical properties of the SPT phases implied by the BR braiding statistics.}

\bigskip

\bigskip
\bigskip

\bigskip

          \bigskip
\setcounter{equation}{0}
\section{Part 1: Particle-loops braiding}\label{Braiding}

In this part, we discuss the links which show up in the particle-loops braiding. In section 1.1, we introduce the link homotopy and link groups \cite{milnor1954link} from a physical point of view. In section 1.2, we discuss the linking numbers associated with the links. In particular, we review the geometric meaning of Hopf linking number and Milnor's triple linking number.

\subsection{1.1 Link homotopy and link groups}

In this section, we introduce the link groups in the context of particle-loops braiding statistics in Abelian discrete gauge theories. Given the loops $m_{a}$ for $a\!=\!1,2,\!\dots,M$ in space, where each of them is of certain flavor. Suppose that a particle is initialized a base point $x_o$, we are interested in the possible braiding trajectories with the same braiding statistics. Quantum mechanically, the trajectory can be continuously deformed without altering the resulting observable statistics. Under the deformation, it can even cross with itself since the point of contact corresponds to the particle positions at different time which do not interact with each other. However, it cannot cross with the loops or otherwise the Aharonov-Bohm effect can change the resulting braiding statistics. Therefore, we are interested in the possible trajectories up to such continuous deformation, called \emph{homotopy}. Let $\mathfrak{m}\!=\!\cup_a m_a\!$ be the region occupied by the loops, the particle can only braid through the   loops complement $\mathfrak{m}^c$. Hence the collection of all possible trajectories is precisely the fundamental group $\pi_1(\mathfrak{m}^c\!\!,x_o)$ of  $\mathfrak{m}^c$. Given the base point, let $b_a$ be the closed path in winding the particle once around $m_a$. By the van Kampen theorem \cite{AlTopo, Knot}, the fundamental group is generated by these closed paths, that is, the fundamental group can be written as
\begin{align}
\pi_1(\mathfrak{m}^c\!\!,x_o)=\langle b_{1},b_{2},\dots,b_{M} |R\rangle\,,
\end{align}
where $R$ is a set of group relations with $R=\emptyset$ iff the loops are unlinked \cite{Knot}. Hence any trajectory can be written as a word in letters $b_a$'s. Note that carrying the particle along $b_a$ is represented as a quantum operator $\mathsf{B}_a$, so any braiding process is a sequential operation in $\mathsf{B}_a$'s as well as their inverses. In such  a construction, although the particle can come back to the base point multiple times in the whole process, the trajectory can always be deformed such that it does not get to the base point during the intermediate period. Generally, one can deform the trajectory such that there is no self intersection. In such a picture, each trajectory $f$ in the fundamental group together with the loops can be viewed as a link,
\begin{align}
L\!=\!(m_{i_1},\dots,m_{i_M},f)\,.
\end{align}
Suppose that the loops are allowed to deform physically.
Recall that the particle-loop braiding within a flavor must yield an Abelian braiding phase. Assume that particle-loop braiding statistics for distinct flavors is allowed to be non-Abelian. Consequently, a loop of certain flavor can host unlocalized particle of other flavors \cite{PRESKILL199050,PhysRevLett.64.1632,ALFORD1992251,BUCHER19923}. For two loops with distinct flavors, one loop may carry unlocalized particle with flavor of the other loop, and hence they can have mutual Aharonov-Bohm interaction. On the contrary, two loops with the same flavor can never interact through Aharonov-Bohm interaction. With such physical picture, the loops $m_{a}$'s do not self-interact but they may have mutual-interaction. In such case, each loop can cross itself but not other loops. Therefore, we have a deformation for the whole link such that each link component in $L$ can cross itself, but no two of them can cross each other. Such link homotopy defines an equivalent relation "$\sim$" in the fundamental group, that is,
\begin{align}
f_1\sim f_2\textrm{\,\,\,\,\,\,\,\,if\,\,\,\,\,\,\,\,}L_1 \textrm{\,is link homotopic to\,} L_2\,.
\end{align}
The collection of physically equivalent trajectories in the fundamental group forms the \emph{link group} of the loops \cite{milnor1954link},
\begin{align}
\mathcal{G}(\mathfrak{m})=\pi_1(\mathfrak{m}^c,x_o)/\sim\,.
\end{align}
Hence, the link group describes the physically distinct braiding trajectories in the particle-loops braiding.
Generally, two links produce the same braiding phase if they can be deformed into one another by a  link homotopy.

\subsection{1.2 Linking numbers}

In the main text, we are interested in the 2-component and 3-component links which correspond to the particle-loop and particle-loop-loop braiding respectively. These links are classified by some linking numbers. More precisely, any link with two components $\gamma^i$ and $\gamma^j$ is determined by the Hopf linking number $\mathfrak{L}(\gamma^i\!,\!\gamma^j)$; any link with three components $\gamma^i$, $\gamma^j$ and $\gamma^k$ is determined by the three mutual Hopf linking numbers and the Milnor's triple linking number $\bar{\mu}(\gamma^i,\gamma^j,\gamma^k)$. In the following, we briefly review the geometric definition of Hopf linking number and Milnor's triple linking number.

\subsubsection{\,\,\,\,1.2.1 Hopf linking number}

Here, we review the Hopf linking number. Consider two closed curves $\gamma^i$ and $\gamma^j$. Let $\mathsf{S}^i$ and $\mathsf{S}^j$ be the corresponding Seifert surfaces. Associated with the intersection of $\gamma^i$ with $\mathsf{S}^j$, we have a sign determined by the direction of the path and the surface normal at the point. Let $I_{ij}$ be the sum of  signed intersections of $\gamma^i$ and $\mathsf{S}^j$, the Hopf linking number is
\begin{align}
\mathfrak{L}(\gamma^i,\gamma^j)=I_{ij}\,,
\end{align}
which is symmetric about $i$ and $j$. Note that traversing along any link component $N$ times scales $\mathfrak{L}$ by $N$. Also, changing the orientation of any link component flips the sign of $\mathfrak{L}$.

\subsubsection{\,\,\,\,1.2.2 Milnor's triple linking number}\label{geo}

Here, we introduce the geometric meaning and some geometric
properties of the Milnor's triple linking number. Consider the three closed curves $\gamma^i$, $\gamma^j$ and $\gamma^k$. The Milnor's triple linking number $\bar{\mu}$ can be written in terms of two geometric quantities $t_{ijk}$ and $m_{ijk}$ \cite{mellor2003geometric}. The first quantity is defined through the simultaneous intersection points of the Seifert surfaces $\mathsf{S}^i$, $\mathsf{S}^j$ and $\mathsf{S}^k$ of the three closed curves. Associated with each intersection point, there is a sign given by the orientation of the normals at the point. The quantity $t_{ijk}$ is the sum of the signed intersections of $\mathsf{S}^i$, $\mathsf{S}^j$ and $\mathsf{S}^k$. The second quantity is defined through the observations in traveling along the three closed curves. Imagine traveling once along $\gamma^k$ from a starting point. The path will sequentially intersect the surfaces $\mathsf{S}^i$ and $\mathsf{S}^j$. Associate each intersection with a sign determined by the direction of the path and the surface normal at the point. In addition, for each occurrence of $\mathsf{S}^i$ after $\mathsf{S}^j$, we can define a sign given by the product of the signed intersections at $\mathsf{S}^i$ and $\mathsf{S}^j$. Let $e_{ijk}$ be the sum of signed occurrences of $\mathsf{S}^i$ after $\mathsf{S}^j$ along $\gamma^k$. Similarly, we have $e_{jki}$ and $e_{kij}$. Note that upon changing the starting point on the three closed curves or interchanging any two indices, $e_{ijk}, e_{jki}$ and $e_{kij}$ are changed by some integral multiple of $\Delta$, where $\Delta$ is the greatest common divisor (gcd) of the mutual Hopf linking numbers of the three curves. The geometric quantity $m_{ijk}$ is given by
\begin{align}
m_{ijk}=e_{ijk}+e_{jki}+e_{kij}\text{ mod }\Delta.
\end{align}
Generally, the Milnor's triple linking number $\bar{\mu}$ of the three closed curves $\gamma^i$, $\gamma^j$ and $\gamma^k$ can be written as
\begin{align}\label{MTL}
\bar{\mu}(\gamma^i,\gamma^j,\gamma^k)=t_{ijk}-m_{ijk}\,\,\,\textrm{mod}\,\Delta\,.
\end{align}
In particular, if the three closed curves are mutually unlinked, which is the case we focused in the main text, we have $\Delta\!=\!0$ and $\bar{\mu}$ is uniquely defined and is an invariant under link homotopy. The Milnor's triple linking number satisfies some interesting geometric properties \cite{milnor1954link}. First, traversing along any link component $N$ times scales $\bar{\mu}$ by $N$. Second, flipping the orientation of any link component changes the sign of $\bar{\mu}$. Moreover, interchanging any two components of the link also flips the sign of $\bar{\mu}$. These properties are useful in the derivation of the geometric properties of the BR braiding statistics in the main text.

\section{Part 2: Topological quantum field theories}

In this part, we supplement technical details for the TQFTs section of the main text. In section 2.1, we introduce the BF theories with an $AAB$ topological term. In section 2.2, we discuss the geometric meaning of the response under sources.

\subsection{2.1 $BF$ theories with an $AAB$ term}

In this section, we discuss the properties of the BF theories with an $AAB$ term. First, we write down the topological actions. Second, we show their variations under gauge transformation. Third, we derive the constraints of the $AAB$ term. Fourth, we derive the gauge invariant braiding process.

\subsubsection{\,\,\,\,2.1.1 Topological actions}\label{ta}

In the main text, we are primarily interested in the BF theories with an
$AAB$ term. For concreteness, we write down the topological actions here. We pick some distinct $i,j,k\in \mathcal{F}$ for the $A^iA^jB^k$ term and focus on the three involved components $\Z_{N_i}\!\times\!\Z_{N_j}\!\times\!\Z_{N_k}$ of the gauge group $G$. Unless otherwise specified, there is no implicit sum for repeated indices. Let the flavor index $\imath\!=\!i,j,k$ and denote $n=N_iN_jN_k$. Then, the action $S$ is the sum of two parts $S_{\rm BF}$ and $S_{\rm AAB}$, that is
\begin{align}\label{ATQFT}
S=\int\sum_\imath\frac{N_\imath}{2\pi} B^\imath dA^\imath+\,\frac{n c_{ij,k} }{(2\pi)^3} A^i A^j B^k\,,\,\,\,\,\,\,\,
\end{align}
where the 1-form $A^{\imath}$ and 2-form $B^{\imath}$ are compact U(1) fields with the Dirac quantization $\oint\!dA^{\imath}\!\in\!2\pi\Z$ and $\oint\!dB^{\imath}\!\in\!2\pi\Z$. Let $a\!=\!i,j$, the fields $\!B^a\!$ and $\!A^k\!$ serve as the Lagrange multipliers which enforce the gauge fields $A^a\!$ and $B^k\!$ to be closed, that is $dA^{a}\!=\!0$ and $dB^{k}\!=\!0$, with cyclic Wilson integrals $\oint\!A^a\!\in\!\frac{2\pi}{N_a}\Z_{N_a}\!$ and $\oint\! B^k\!\in\!\frac{2\pi}{N_k}\Z_{N_k}\!$. In the presence of external sources, the action acquires an extra term
\begin{align}\label{ASource}
&\,\,\,\,S_{\rm s}
=-\!\int\!\sum_{a}\big(J^a\!A^a\!+\!\Sigma^a\mathcal{B}^a\big)\!+\!\Sigma^kB^k\!+\!J^k\mathcal{A}^k,
\end{align}
where the 3-form $J^{\imath}$ and 2-form $\Sigma^{\imath}$ with $dJ^{\imath}\!=\!0$ and $d\Sigma^{\imath}\!=\!0$ are respectively the conserved particle and loop sources. In particular, $\Sigma^{a}$ and $J^{k}$ are respectively coupled to the modified lagrange multipliers $\mathcal{B}^a$ and $\mathcal{A}^k$ which are given by
\begin{align}
&\,\,\,\,\,\,\,\,\mathcal{B}^a=B^a\!-\!\sum_{b}\!\frac{nc_{ab,k}}{2(2\pi)^2 N_a}\!(A^bd^{-1}\!B^k\!-\!d^{-1}\!A^b\,B^k)\,,\\
&\,\,\,\,\,\,\,\,\mathcal{A}^k=A^k\!-\!\sum_{ab}\!\frac{nc_{ab,k}}{2(2\pi)^2 N_k}\!A^{a}d^{-1}\!A^b\,,
\end{align}
where the indices $a,b$ take only on the particular values $i,j$. Note that the 3-form $J^{\imath}$ and 2-form $\Sigma^{\imath}$ sources here represent the dual of the usual 1-form particle current and 2-form loop current. In such convention, the usual minimal coupling can be simplified as wedge product of sources and gauge fields. As a side remark, the differential forms $A^\imath$ and $B^\imath$ can be written in component form as $A^\imath\!\!=\!\!A^\imath_\mu dx^\mu$ and $B^\imath\!\!=\!\!\frac{1}{2!}B^\imath_{\mu\nu} dx^{\mu\nu}$, where the repeated spacetime greek indices are summed over. In such notation, we have \small$S_{\rm BF}\!\!=\!\!\int\sum_\imath\frac{N_\imath}{4\pi} \epsilon^{\mu\nu\rho\sigma}\! B^\imath_{\mu\nu}\partial_\rho A^\imath_\sigma d^4x$\normalsize and \small$S_{\rm AAB}\!\!=\!\!\int \frac{1}{2!}\frac{nc_{ij,k}}{(2\pi)^3}\epsilon^{\mu\nu\rho\sigma}\!A_\mu^iA_\nu^jB_{\rho\sigma}^kd^4x$\normalsize, where $\epsilon^{\mu\nu\rho\sigma}$ is the Levi-Civita symbol. Next, we are going to study the physical properties of such type of TQFTs.

\subsubsection{\,\,\,\,2.1.2 Gauge transformation}\label{gau}

Here, we show details of the gauge transformation of the BF theories with an
$AAB$ topological term. We are going to illustrate the variations of the actions $S$, $S_{\rm AAB}$ and $S_{\rm s}$ in Eq.\ (\ref{ATQFT}) and (\ref{ASource}) separately under the gauge transformation,
\begin{align}
A^a&\rightarrow A^a+d\alpha^a, \quad B^a\rightarrow  B^a+d\beta^a+\mathcal{X}^a\,,\\
B^k&\rightarrow B^k+d\beta^k, \quad A^k\rightarrow A^k+d\alpha^k+\mathcal{X}^k\,,
\end{align}
where the 0-form $\alpha^{\imath}$ and 1-form $\beta^{\imath}$ are compact U(1) gauge parameters with the windings $\oint d\alpha^{\imath}\!\in\!2\pi\Z$ and $\oint d\beta^{\imath}\!\in2\!\pi\Z$. Note that the Lagrange multipliers $B^a$ and $A^k$ transform with extra twists $\mathcal{X}^a$ and $\mathcal{X}^k$, which are given by
\begin{align}
\quad\,\,\mathcal{X}^{a}&=-\sum_{b}\frac{n c_{ab,k}}{(2\pi)^2 N_a}(\alpha^{b}B^k-A^{b}\beta^k+\alpha^{b}d\beta^k)\,,\\
\quad\,\,\mathcal{X}^k&=-\sum_{ab}\frac{n c_{ab,k}}{(2\pi)^2 N_k}(\alpha^a A^{b}+\frac{1}{2}\alpha^a d\alpha^b)\,,
\end{align}
where the indices $a,b\!=\!i,j$. The compatibility of the gauge transformation with the Dirac quantization of $B^a$ and $A^k$ gives $\oint d\mathcal{X}^{\imath}\!\in\!2\pi\Z$ as a requirement. Equivalently, the gauge transformation above can be rewritten as
\begin{align}
A^a&\rightarrow A^a+d\alpha^a, \quad \mathcal{B}^a\rightarrow  \mathcal{B}^a+d\beta^a+\mathcal{Y}^a\,,\\
B^k&\rightarrow B^k+d\beta^k, \quad \mathcal{A}^k\rightarrow \mathcal{A}^k+d\alpha^k+\mathcal{Y}^k\,,
\end{align}
where the modified Lagrange multipliers $\mathcal{B}^a$ and $\mathcal{A}^k$ transform with extra twists $\mathcal{Y}^a$ and $\mathcal{Y}^k$, which are defined by
\begin{align}
\quad\,\,\mathcal{Y}^{a}&=-\sum_{a}\frac{nc_{ab,k}}{2(2\pi)^2 N_a}d(\alpha^bd^{-1}\!B^k-\beta^kd^{-1}\!A^b)\,,\\
\quad\,\,\mathcal{Y}^k&=-\sum_{ab}\frac{nc_{ab,k}}{2(2\pi)^2 N_k}d(\alpha^ad^{-1}\!A^b)\,.
\end{align}
Notice that, unlike $\mathcal{X}^a$ and $\mathcal{X}^k$, the twists $\mathcal{Y}^a$ and $\mathcal{Y}^k$ are like total derivatives, hence the modified lagrange multiplier fields $\mathcal{B}^a$ and $\mathcal{A}^k$ transform like ordinary gauge fields. However, since $\mathcal{Y}^a$ and $\mathcal{Y}^k$ involve derivatives of non-local terms, they are not exactly total derivatives. Hence, $\mathcal{B}^a$ and $\mathcal{A}^k$ generally do not transform with a pure gauge.

The details of the variations of the actions $S$, $S_{\rm AAB}$ and $S_{\rm s}$ under the gauge transformation are given as follows. As for the total action $S$, it transforms as
\begin{align}
S\rightarrow S+\Delta S^{(1)}+\Delta S^{(2)}+\Delta S^{(3)},
\end{align}
where the changes are respectively the 1st, 2nd and 3rd order changes of $S$ in the gauge parameters, and
\begin{align}
\Delta S^{(1)}
=&\int \sum_a \frac{N_a}{2\pi}d\beta^adA^a+\frac{N_k}{2\pi}d\alpha^kdB^k\\
&\,+\sum_{ab}-\frac{nc_{ab,k}}{(2\pi)^{3}}\big(d(\alpha^aA^bB^k)+\frac{1}{2}d(A^aA^b\beta^k)\big)\,,\nonumber\\
\Delta S^{(2)}
=&\int \sum_{ab}-\frac{nc_{ab,k}}{(2\pi)^{3}}d\alpha^ad(\alpha^bB^k-A^b\beta^k)\nonumber\\
&\,+\sum_{ab}-\frac{nc_{ab,k}}{(2\pi)^{3}}d(\alpha^aA^b)d\beta^k\,,\nonumber\\
\Delta S^{(3)}
=&\int \sum_{ab}-\frac{nc_{ab,k}}{(2\pi)^{3}}d\alpha^ad\alpha^bd\beta^k\,,\nonumber
\end{align}
which are surface terms. Integrating out the Lagrange multipliers $B^{a}$ and $A^k$ reduces
$S$ to $S_{\rm AAB}$.
Now, under the gauge transformation of the remaining fields $A^{a}$ and $B^k$,
\begin{align}\label{SAABc}
S_{\rm AAB}\rightarrow S_{\rm AAB}+\Delta S_{\rm AAB}^{(1)}+\Delta S_{\rm AAB}^{(2)}+\Delta S_{\rm AAB}^{(3)}\,,
\end{align}
where the changes are respectively the 1st, 2nd and 3rd order changes of $S_{\rm
  AAB}$
in the gauge parameters, and
\begin{align}\label{quan}
\Delta S_{\rm AAB}^{(1)}
=&\int \sum_{ab}\frac{nc_{ab,k}}{(2\pi)^{3}}(d\alpha^aA^bB^k+\frac{1}{2}A^aA^bd\beta^k)\,,\\
\Delta S_{\rm AAB}^{(2)}
=&\int \sum_{ab} \frac{nc_{ab,k}}{(2\pi)^{3}}(\frac{1}{2}d\alpha^ad\alpha^bB^k+d\alpha^aA^bd\beta^k)\,,\nonumber\\
\Delta S_{\rm AAB}^{(3)}
=&\int \sum_{ab} \frac{nc_{ab,k}}{(2\pi)^{3}}\frac{1}{2}d\alpha^ad\alpha^bd\beta^k\,.\nonumber
\end{align}
Finally, consider the total action $S$ coupled to the sources described by
$S_{\rm s}$.
Under the gauge transformation,
\begin{align}\label{sch1}
S^{}_{\rm s}\rightarrow S^{}_{\rm s} + \Delta S_{\rm s}^{\rm pure}+\Delta S_{\rm s}^{\rm twist}\,,
\end{align}
where the 1st term is the change due to the gauge parameters and the 2nd term is the change due to the twists $\mathcal{Y}^a$ and $\mathcal{Y}^k$, and the two contribution are respectively given by
\begin{align}\label{sch2}
\Delta S_{\rm s}^{\rm pure}
&=\int\sum_{a}-(J^ad\alpha^a+\Sigma^ad\beta^a)-(\Sigma^kd\beta^k+J^kd\alpha^k)\,,\nonumber\\
\Delta S_{\rm s}^{\rm twist}
&=\int\sum_{ab}\frac{nc_{ab,k}}{2(2\pi)^2 N_a}\Sigma^ad(\alpha^bd^{-1}\!B^k-\beta^kd^{-1}\!A^b)\nonumber\\
&\quad\,\,+\sum_{ab}\frac{nc_{ab,k}}{2(2\pi)^2 N_k}J^kd(\alpha^ad^{-1}\!A^b)\,.
\end{align}
The gauge transformation shown here are useful in the field theoretical derivation of properties of the BR braiding statistics in discrete gauge theories.

\subsubsection{\,\,\,\,2.1.3 Constraints on the $S_{\rm AAB}$ topological term}\label{inv1}

Here, we discuss the constraints on the topological term $S_{\rm AAB}$ in Eq.\ (\ref{ATQFT}). More precisely, we discuss the properties of the coefficient $c_{ij,k}$. In the main text, we have seen
\begin{align}
c_{ij,k}=-c_{ji,k}\,,
\end{align}
due to antisymmetry of the $A^iA^jB^k$ term upon interchanging $A^i$ and $A^j$. In addition, we have also shown that
\begin{align}
c_{ij,k}=0\,,\quad\textrm{if}\,\,i,j,k\,\,\textrm{are not distinct,}
\end{align}
for proper cyclic fusion structure of the discrete gauge theories. Here, we are going to show that $c_{ij,k}$ is quantized as
\begin{align}\label{Aquan}
c_{ij,k}=\frac{2\pi k_{ij,k}}{N_{ijk}},
\end{align}
where $k_{ij,k}$ is an integer and the number $N_{ijk}$ is the gcd of $N_i, N_j$ and $N_k$. Note that all properties of the coefficient $c_{ij,k}$ naturally propagate to $k_{ij,k}$, that is, $k_{ji,k}\!=\!-k_{ij,k}$ and $k_{ij,k}$ vanishes if any of the indices are identical.

We now derive the quantization of $c_{ij,k}$. To be precise, we consider $S$ without source. Let $a=i,j$. After integrating out the Lagrange multipliers $B^a$ and $A^k$, the action $S$ reduces to $S_{\rm AAB}$, where the fields $A^{a}$ and $B^k$ are set to be closed with $\oint A^a\!\in\!\frac{2\pi}{N_a}\Z_{N_a}$ and $\oint B^k\!\in\!\frac{2\pi}{N_k}\Z_{N_k}$. Note that any legitimate TQFT is well defined on arbitrary orientable closed manifold. In particular, for the moment, consider putting the TQFT on the spacetime manifold $\MM=\mathbb{S}^1\times\mathbb{S}^1\times\mathbb{S}^2$. Under the large gauge transformation $A^a\rightarrow A^a+d\alpha^a$ and $B^k\rightarrow B^k+d\beta^k$, the topological term $S_{\rm AAB}$ is changed as in Eq.\ (\ref{SAABc}), where the 1st, 2nd and 3rd order changes are respectively given by $\Delta S_{\rm AAB}^{(1)}$, $\Delta S_{\rm AAB}^{(2)}$ and $\Delta S_{\rm AAB}^{(3)}$ in Eq.\ (\ref{quan}). Suppose that
\begin{align}
&\oint_{\mathbb{S}^1} d\alpha^a=2\pi p^a\,\,\,\,\,\,\mbox{and}\,\,\,\,\,\,\oint_{\mathbb{S}^2} d\beta^k=2\pi p^k\,,\\
&\oint_{\mathbb{S}^1} A^a=\frac{2\pi q^a}{N_a}\,\,\,\,\,\,\mbox{and}\,\,\,\,\,\,\oint_{\mathbb{S}^2} B^k=\frac{2\pi q^k}{N_k}\,,
\end{align}
where $p^a,p^k,q^a,q^k$ are some integers. In particular,  $(d\alpha^i, A^i)$ and $(d\alpha^j,A^j)$ wind around the first and the second copy of $\mathbb{S}^1$ in $\MM$ respectively, and $(d\beta^k,B^k)$ winds only around the $\mathbb{S}^2$ in $\MM$. Then each integral in Eq.\ (\ref{quan}) can be calculated explicitly,
\begin{align}
&\int_{\MM} \!\!d\alpha^aA^bB^k=\oint_{\mathbb{S}^1}\!\! d\alpha^a \!\!\oint_{\mathbb{S}^1}\!\! A^b\!\!\oint_{\mathbb{S}^2}\!\! B^k=\frac{(2\pi)^3p^aq^bq^k}{N_bN_k}\,,\\
&\int_{\MM}\!\! A^aA^bd\beta^k=\oint_{\mathbb{S}^1}\!\! A^a \!\!\oint_{\mathbb{S}^1}\!\! A^b \oint_{\mathbb{S}^2}\!\! d\beta^k=\frac{(2\pi)^3q^aq^bp^k}{N_aN_b}\,,\nonumber\\
&\int_{\MM}\!\! d\alpha^ad\alpha^bB^k=\oint_{\mathbb{S}^1}\!\! d\alpha^a\!\! \oint_{\mathbb{S}^1}\!\! d\alpha^b\!\!\oint_{\mathbb{S}^2}\!\! B^k=\frac{(2\pi)^3p^ap^bq^k}{N_k}\,,\nonumber\\
&\int_{\MM}\!\! d\alpha^a A^bd\beta^k=\oint_{\mathbb{S}^1}\!\! d\alpha^a \!\!\oint_{\mathbb{S}^1}\!\! A^b\!\!\oint_{\mathbb{S}^2}\!\! d\beta^k=\frac{(2\pi)^3p^aq^bp^k}{N_b}\,,\nonumber\\
&\int_{\MM}\!\! d\alpha^ad\alpha^bd\beta^k=\oint_{\mathbb{S}^1}\!\! d\alpha^a\!\! \oint_{\mathbb{S}^1}\!\! d\alpha^b\!\! \oint_{\mathbb{S}^2}\!\! d\beta^k=(2\pi)^3p^ap^bp^k\,,\nonumber
\end{align}
where the integral over ${\MM}$ are evaluated along the 1st $\mathbb{S}^1$ and the 2nd $\mathbb{S}^1$ and the last $\mathbb{S}^2$ independently, and $a,b=i,j$. Substituting the results above into Eq.\ (\ref{quan}), we get
\begin{align}
&\Delta S_{\rm AAB}^{(1)}
\!=\!N_{i}c_{ij,k} p^iq^jq^k\!-\!N_{j}c_{ij,k} q^ip^jq^k\!+\!N_{k}c_{ij,k}q^iq^jp^k\,,\nonumber\\
&\Delta S_{\rm AAB}^{(2)}
\!=\!N_{i}N_{j}c_{ij,k} p^i\!p^j\!q^k\!\!-\!\!N_{j}N_{k}c_{ij,k}q^i\!p^j\!p^k\!\!+\!\!N_{i}N_{k}c_{ij,k} p^i\!q^j\!p^k\!\!,\nonumber\\
&\Delta S_{\rm AAB}^{(3)}
\!=\!N_iN_jN_kc_{ij,k}p^ip^jp^k\,.
\end{align}
Since the topological term $S_{\rm AAB}$ is large gauge invariant mod $2\pi$ for arbitrary choice of integers $p^a,p^k,q^a,q^k$, each term above must be quantized to integral multiple value of $2\pi$. Note that the products of the integers $p$'s and $q$'s are still integers. So the quantization of the terms in $\Delta S_{\rm AAB}^{(1)}$ implies that $\!N_{i}c_{ij,k}, N_{j}c_{ij,k}\!$ and $\!N_{k}c_{ij,k}\!\!=\!\!0\,\textrm{mod}\,\,2\pi$. Therefore we have the desired coefficient quantization condition $c_{ij,k}=\frac{2\pi k_{ij,k}}{N_{ijk}}$ in  Eq.\ (\ref{Aquan}). Under such quantization, the 2nd and 3rd order gauge changes $\Delta S_{\rm AAB}^{(2)}$ and $\Delta S_{\rm AAB}^{(3)}$ always take values in integral multiple of $2\pi$, which is automatically compatible to the large gauge invariance.

\subsubsection{\,\,\,\,2.1.4 Gauge invariant braiding process}\label{inv2}

Here, we study the gauge invariant braiding process. Consider a braiding described by the closed world line $\Gamma^{\imath}$ for $e_{\imath}$ and the closed world-sheet $\mathcal{S}^{\imath}$ for $m_{\imath}$ for $\imath\!=\!i,j,k$. They correspond to the conserved sources $J^\imath\!=\!\delta(\Gamma^\imath)$ and $\Sigma^\imath\!=\!\delta(\mathcal{S}^\imath)$, which are delta-forms with supports on the given submanifolds. We will show that the process is gauge invariant iff
\begin{align}
&I_{\rm Hopf}[\Sigma^a,J^k]=\int \Sigma^ad^{-1}J^k=0\,,\,\,\mbox{and}\label{ginvp1}\\
&I_{\rm Hopf}[\Sigma^a,\hat{J}^b]=\int\Sigma^{a} d^{-1}\hat{J}^b=0\,\,\,\,\,\,(a\neq b)\,,\label{ginvp2}
\end{align}
where $\hat{J}^b\!\!=\!\!\delta(\hat{\Gamma}^b)$, for any closed curve $\hat{\Gamma}^b$ on $\mathcal{S}^b$, $a,b\!=\!i,j$. For the first constraint, it means that the particle-loop braiding of $e_k$ with $m_a$ alone is not gauge invariant for $a=i,j$. For the second constraint, its meaning can be extracted by considering different construction of the $\hat{\Gamma}^b$. Take $\hat{\Gamma}^b$ as the world line of a point on the loop $m_b$, it means that particle-loop braiding of such point with $m_a$ is not gauge invariant. Since such argument is true for any choice of point on $m_b$, so loop-loop braiding \cite{baez2007, NovString, ALFORD1992251} of $m_a$ and $m_b$ alone is not gauge invariant for $a\!\neq\! b$. Take $\hat{\Gamma}^b$ as a time slice of $\mathcal{S}^b$, where its trajectory corresponds to the locus of $m_b$, it means that $m_a$ and $m_b$ is always unlinked for $a\!\neq\! b$. In particular, the crossing of $m_a$ and $m_b$, which always changes the linking number, is not allowed in gauge invariant braiding process. Suppose the loops are kept static, the gauge invariant constraints above simply mean that the particle trajectory and the loops $m_i$ and $m_j$ are mutually unlinked circles.

We now derive the gauge invariant braiding processes. For any observable braiding statistics,
the corresponding braiding process described by $S_{\rm s}$ must be gauge invariant. Under the gauge transformation, the source term $S_{\rm s}$ is changed by $\Delta S_{\rm s}^{\rm pure}\!$ and $\Delta S_{\rm s}^{\rm twist}\!$ as in Eq.\ (\ref{sch1}). In terms of the world line $\Gamma^\imath$ and world sheet $\mathcal{S}^\imath$, the gauge changes can be written as
\begin{align}
\Delta S_{\rm s}^{\rm pure}
&=\sum_{\imath}\big(-\int_{\Gamma^\imath} d\alpha^\imath-\int_{\mathcal{S}^\imath} d\beta^\imath\big)\nonumber\\
\Delta S_{\rm s}^{\rm twist}
&=\sum_{ab}\frac{nc_{ab,k}}{2(2\pi)^2 N_a}\int_{\mathcal{S}^a} d(\alpha^bd^{-1}\!B^k-\beta^kd^{-1}\!A^b)\nonumber\\
&\,+\sum_{ab}\frac{nc_{ab,k}}{2(2\pi)^2 N_k}\int_{\Gamma^k}d(\alpha^ad^{-1}\!A^b)\,.
\end{align}
Therefore, both of them have to be vanishing for the gauge invariance. By the Stokes' theorem, since the world line $\Gamma^\imath$ and world sheet $\mathcal{S}^\imath$ are closed, the pure part $\Delta S_{\rm s}^{\rm pure}\!$ is always equal to zero. Similarly, by the Stokes' theorem, the twist part $\Delta S_{\rm s}^{\rm twist}\!$ equals zero iff the integrands are total derivatives over the domains of integration. Since the integrands of the twist part consist of exterior derivatives of non-local terms, they are generally not total derivatives. The integrands of the twist part are total derivatives iff. the non-local terms are globally defined over the domains of integration. It means that $d^{-1}\!B^{k}$ and $d^{-1}\!A^{b}$ are globally defined on $\mathcal{S}^a$ for $a\!\neq\! b$, and $d^{-1}\!A^{b}$ is globally defined on
$\Gamma^k$. Equivalently, $B^{k}$ and $A^{b}$ are exact on $\mathcal{S}^a$ for $a\!\neq\! b$,
and $A^{b}$ is exact on $\Gamma^k$. By the de Rham theorem, for any $p$-form $\omega$ defined on a $n$-dimensional manifold $\Omega$, $\omega$ is exact iff the integral of $\omega$ over any $p$-dimensional closed submanifold is zero. Therefore
\begin{align}\label{con}
\,\,\,\,\,\,
\int_{\mathcal{S}^a} B^k=0\,,\,\,
\int_{\hat{\Gamma}^a} A^b=0\,\,\,\,(a\neq b)\,,\,\,
\int_{\Gamma^k} A^b=0\,,
\end{align}
for any closed curve $\hat{\Gamma}^a$ on $\mathcal{S}^a$. With the source term $S_{\rm s}$, the Lagrange multiplier fields $B^a$ and $A^k$ enforce the equations $\Sigma^a\!=\!\frac{N_a}{2\pi}dA^{a}$ and $J^k\!=\!\frac{N_k}{2\pi}dB^{k}$. In terms of these sources, the third constraint above becomes redundant, and Eq.\ (\ref{ginvp1}) and (\ref{ginvp2}) follow immediately from Eq.\ (\ref{con}).

\subsection{2.2 Geometric meaning of the effective action}\label{thm}

In this section, we discuss the geometric interpretation of the effective action. More precisely, we study the geometric meaning of the effective action by dimension reduction technique. First, we write down the effective action under the source term $S_s$. Such effective action is an integral over spacetime. Second, we introduce the dimension reduction method. Third, we illustrate the reduction of the effective action. Under the dimension reduction, the effective action reduces to an integral over space only. Fourth, we show that the spatial integral yields the Hopf linking number and the Milnor's triple linking number.

\subsubsection{\,\,\,\,2.2.1 Effective action}\label{thm}

Here, we write down the effective action $S_{\rm eff}$ of the TQFTs. Under the external sources $J^\imath$ and $\Sigma^\imath$ for particles and loops, the action acquires an addition source term $S_{\rm s}$. With the extra source term $S_{\rm s}$, the Lagrange multipliers $B^a$ and $A^k$ impose instead the set of constraints $\Sigma^a\!=\!\frac{N_a}{2\pi}dA^{a}$ and $J^k\!=\!\frac{N_k}{2\pi}dB^{k}$, where $a\!=\!i,j$. After integrating out the Lagrange multipliers, the action $S\!+\!S_s$ is turned into the effective action
\begin{align}
S_{\rm eff}=\sum_\imath\frac{2\pi}{N_{\imath}}I_{\rm Hopf}[\Sigma^\imath\!,J^\imath]
+\frac{2\pi k_{ijk}}{N_{ijk}}I_{\rm BR}[\Sigma^i\!,\Sigma^j\!,J^k]\,,
\end{align}
where the index $\imath\!=\!i,j,k$. In the expression, $I_{\rm Hopf}$, which arises from  $S_{\rm BF}$, is given by the spacetime integral
\begin{align}\label{SHOPFM}
I_{\rm Hopf}[\Sigma^\imath\!,J^\imath]=\int_{\MM} \Sigma^\imath d^{-1}\!J^\imath\,,
\end{align}
whereas $I_{\rm BR}$, which arises from the  $AAB$ topological term $S_{\rm AAB}$, is defined by the spacetime integral
\begin{align}\label{SBRM}
I_{\rm BR}[\Sigma^i\!,\Sigma^j\!,&J^k]
=\int_{\MM}  d^{-1}\Sigma^id^{-1}\Sigma^jd^{-1}J^k \nonumber\\
&- \frac{1}{2}\Sigma^i(d^{-1}\Sigma^jd^{-2}J^k-d^{-1}J^kd^{-2}\Sigma^j) \nonumber\\
&- \frac{1}{2}\Sigma^j(d^{-1}J^kd^{-2}\Sigma^i-d^{-1}\Sigma^id^{-2}J^k)\nonumber\\
&- \frac{1}{2}J^k(d^{-1}\Sigma^id^{-2}\Sigma^j-d^{-1}\Sigma^jd^{-2}\Sigma^i)\,.
\end{align}
The remaining parts of this section is devoted to studying the geometric meaning of these two spacetime linking invariants $I_{\rm Hopf}$ and $I_{\rm BR}$. We will see that while the first integral $I_{\rm Hopf}$ counts the Hopf linking number between $m_\imath$ and $\gamma_{e_\imath}$, the second integral $I_{\rm BR}$ counts the Milnor's triple linking number between $m_i$, $m_j$ and $\gamma_{e_k}$.

\subsubsection{\,\,\,\,2.2.2 Dimension reduction method}\label{DR}

Here, we discuss the dimension reduction of the integral of differential forms over the spacetime $\MM=M\!\times\!T$. Since the coefficients of differential forms have to be taken care, we denote $\mu$ as the spacetime index and $i,j$ as the spatial indices without confusion here.  Also, implicit sum is imposed on repeated indices. Let $\omega$ be a $p$-form and $\lambda$ be a $q$-form on $\MM$, where $p+q$ equals the dimension of $\MM$. We are interested in the dimension reduction of integral of the form $\mathfrak{I}=\int_{\MM}\lambda\omega$. The integral $\mathfrak{I}$ is said to be reducible if $\omega$ is static with temporal gauge, and $\lambda$ is in cotemporal gauge [Eq.\ (\ref{type1}) and (\ref{type2})]. We are going to show that if $\mathfrak{I}$ is reducible, then the spacetime integral $\mathfrak{I}$  can be written as a spatial integral as
\begin{align}
\mathfrak{I}=\int_{\MM}\lambda\omega=\int_{M}\lambda_R\omega_R\,,
\end{align}
where the $(p-1)$-form $\omega_R$ and the $q$-form $\lambda_R$ are respectively the reduction of $\omega$ and $\lambda$ defined on $M$ [Eq.\ (\ref{re1}) and (\ref{re2})].


We first define the two types of differential forms $\omega$ and $\lambda$. Let $\omega_{\mu_1\dots\mu_p}$ and $\lambda_{\mu_1\dots\mu_q}$  be the anti-symmetrized coefficients of $\omega$ and $\lambda$ respectively. The $p$-form $\omega$ is static if its coefficient is time independent, and it is in temporal gauge if its temporal component $\!\omega_{0i_1\dots i_{p-1}}\!$ vanishes. For static $\omega$ in temporal gauge,
\begin{align}\label{type1}
\omega=\omega_{i_1\dots i_p}dx^{i_1\dots i_p}\,,
\end{align}
where $\omega_{i_1\dots i_p}$ is time independent. The $q$-form $\lambda$ is said to be in cotemporal gauge if its spatial component $\lambda_{i_1\dots i_p}$ vanishes,
\begin{align}\label{type2}
\lambda=\lambda_{0j_1\dots j_{q-1}}dx^{0j_1\dots j_{q-1}}\,.
\end{align}
By considering the exterior derivatives of the equations above, we see that the defining  properties of $\omega$ and $\lambda$ are closed under $d$. Since the inverse operation $d^{-1}\!$ can in principle introduce extra gauge degrees of freedom, the properties of the two types of forms may not be preserved under such operation. Nevertheless, by comparing the coefficient in the identity $dd^{-1}\omega\!=\!\omega$, we see that $d^{-1}\omega$ is guaranteed to be static.


We now define the reductions $\omega_R$ and $\lambda_R$ for $\omega$ and $\lambda$  separately. For the static $p$-form $\omega$ in temporal gauge, its reduction $\omega_R$ is the restriction of $\omega$ on the time slice $t\!=\!t_0\!\in\!T$,
\begin{align}\label{re1}
\omega_R=\omega|_{t=t_0}=\omega_{i_1\dots i_p}dx^{i_1\dots i_p}\,,
\end{align}
which is defined on $M$. Since the restriction on a time slice set $dx^0\!=\!0$, such reduction is still well-defined if $\omega$ is not in temporal gauge. For the $q$-form $\lambda$ in cotemporal gauge, its reduction $\lambda_R$ is the integral of $\lambda$ along the time direction,
\begin{align}\label{re2}
\lambda_R=\int_{T}\lambda=\Big(\!\int_{T}\lambda_{j_1\dots j_{q-1}}dx^{0}\Big)dx^{j_1\dots j_{q-1}}\,,
\end{align}
which is defined on $M$. Meanwhile, we show the commutation properties of reduction with exterior derivative, that is, the reduction of $\omega$ commutes with $d$ and $d^{-1}$, and the reduction of $\lambda$ commutes with $d$. Since $\omega$ is static and $\omega_R$ is defined on the time slice, $(d\omega)_R\!=\!\PP_{i_1}\omega_{i_2\dots i_{p+1}}dx^{i_1i_2\dots i_{p+1}}\!=\!d(\omega_R)$. Since $d^{-1}\omega$ is static and $(d^{-1}\omega)_R$ is defined on the time slice, $(d^{-1}\omega)_R\!\!=\!\!d^{-1}d(d^{-1}\omega)_R\!\!=\!\!d^{-1}(\omega_R)$. Besides, since $d\lambda$ is in temporal gauge and $\lambda_R$ is defined on the time slice, we have $(d\lambda)_R\!\!=\!\!\PP_{j_1}\!(\!\int_{T}\!\lambda_{j_2\dots j_{q-1}}dx^{0}\!)dx^{j_1j_2\dots j_{q}}\!\!=\!\!d(\lambda_R).$  Note that $d^{-1}\!\lambda\!$ may not be in temporal gauge, its reduction is generally not well defined, so reduction of $\lambda$ may not commute with $d^{-1}$.

We now derive the result for dimension reduction. Consider the spacetime integral $\mathfrak{I}\!=\!\int_{\MM}\lambda\omega$. Note that in the integrand of $\mathfrak{I}$, if $\omega$ is in temporal gauge, then $\lambda$ can be taken to be in cotemperal gauge. Conversely, if $\lambda$ is in cotemperal gauge, then $\omega$ can be taken to be in temporal gauge. The integral $\mathfrak{I}$ is said to be reducible if $\omega$ is static with temporal gauge and $\lambda$ is in cotemporal gauge. If the integral $\mathfrak{I}$ is reducible, then
\begin{align}
&\int_{\MM}\!\!\lambda\omega
=\!\!\int_{M\times T}\!\!\!\!\lambda_{0j_1\dots j_{q-1}}\omega_{i_1\dots i_p}dx^{0j_1\dots j_{q-1}i_1\dots i_p}\\
&=\!\!\int_{M}\!\!\Big(\!\!\int_{T}\!\!\lambda_{j_1\dots j_{q-1}}dx^{0}\Big)\omega_{i_1\dots i_p}dx^{j_1\dots j_{q-1}i_1\dots i_p}\!=\!\!\int_{M}\!\!\lambda_R\omega_R\,.\nonumber
\end{align}
Hence the spacetime integral $\mathfrak{I}$ can be written as a spatial integral if it is reducible. Such reduction is useful in understanding the spacetime linking invariants which show up in the TQFTs of the main text.

\subsubsection{\,\,\,\,2.2.3 Reduction of $I_{\rm Hopf}$ and $I_{\rm BR}$}\label{dm}

Here, we express the spacetime integrals $I_{\rm Hopf}$ and $I_{\rm BR}$ as spatial integrals by dimension reduction. We are going to show that, if the loops are fixed to be static with vanishing background particle density, then the spacetime integral $I_{\rm Hopf}$ shown in Eq.\ (\ref{SHOPFM}) reduces to the following spatial integral
\begin{align}\label{SHOPFS}
I_{\rm Hopf}[\Sigma^\imath\!,J^\imath]=\int_{M} \Sigma^\imath_R d^{-1}\!J^\imath_R\,,
\end{align}
where $\imath\!=\!i,j,k$, in addition, the spacetime integral $I_{\rm BR}$ shown in Eq.\ (\ref{SBRM}) can be written as the following spatial integral
\begin{align}\label{SBRS}
\!I_{\rm BR}[\Sigma^i\!,\Sigma^j\!,&J^k]
=\int_{M}  d^{-1}\Sigma^i_Rd^{-1}\Sigma^j_Rd^{-1}J^k_R\nonumber\\
&- \frac{1}{2}\Sigma^i_R(d^{-1}\Sigma^j_Rd^{-2}J^k_R-d^{-1}J^k_Rd^{-2}\Sigma^j_R)\nonumber\\
&- \frac{1}{2}\Sigma^j_R(d^{-1}J^k_Rd^{-2}\Sigma^i_R-d^{-1}\Sigma^i_Rd^{-2}J^k_R)\nonumber\\
&-\frac{1}{2}J^k_R(d^{-1}\Sigma^i_Rd^{-2}\Sigma^j_R-d^{-1}\Sigma^j_Rd^{-2}\Sigma^i_R)\,,
\end{align}
where $\Sigma^{\imath}_R\!\!=\!\!\Sigma^{\imath}|_{t=t_0}$ and $J^\imath_R\!\!=\!\!\int_{T}\!J^\imath$. Besides, given the sources $\Sigma^\imath\!\!=\!\!\delta(\mathcal{S}^\imath)$ and $J^\imath\!\!=\!\!\delta(\Gamma^\imath)$, then $\Sigma^\imath_R\!\!=\!\!\delta(m_\imath)$ and $J^\imath_R\!=\!\delta(\gamma_{e_\imath})$. Physically, the reductions $\Sigma^\imath_R$ and $J^\imath_R$ describe the loop $m_\imath$ and the particle trajectory $\gamma_{e_\imath}$ respectively.



We now derive the reduction of integrals $I_{\rm Hopf}$ and $I_{\rm BR}$. We first write the integrals in terms of some gauge potentials and then perform the dimension reduction. Since the sources $\Sigma^{\imath}$ and $J^\imath$ are conserved, we get $\Sigma^{\imath}\!=\!\frac{1}{2\pi}d\textsf{A}^{\imath}$ and $J^\imath\!=\!\frac{1}{2\pi}d\textsf{B}^\imath$, for some $1$-form $\textsf{A}^{\imath}$ and $2$-form $\textsf{B}^\imath$ where $\imath\!=\!i,j,k$. In term of these gauge potentials, Eq.\ (\ref{SHOPFM}) and (\ref{SBRM}) become
\begin{align}
I_{\rm Hopf}\!=\!\frac{1}{(2\pi)^2}\!\!\int_{\MM}\!\!d\textsf{A}^\imath \textsf{B}^\imath\,,\,\,\,\,\,\,
I_{\rm BR}\!=\!\frac{1}{(2\pi)^3}\!\!\int_{\MM}\!\!\Lambda^{ij}\textsf{B}^k\,,
\end{align}
where the 2-form $\Lambda^{ij}$ is obtained by integration by part and it can be written in terms of the gauge potentials $\textsf{A}^{i}$ and $\textsf{A}^{j}$ as
\begin{align}\label{gam}
\Lambda^{ij}=\sum_{ab}\frac{1}{2}\varepsilon^{ab}\big(2d\textsf{A}^a d^{-1}\!\textsf{A}^b+d^{-1} (d\textsf{A}^a\textsf{A}^b)\big)\,,
\end{align}
where $a,b\!=\!i,j$. By noting that  the parent action $S+S_s$ depends only on $\Sigma^{\imath}$ and $J^\imath$, $I_{\rm Hopf}$ and $I_{\rm BR}$ must be independent on the gauge choices of $\textsf{A}^{\imath}$ and $\textsf{B}^\imath$. For static loops with zero background particle density, we can pick static $\textsf{A}^{\imath}$ in temporal gauge and $\textsf{B}^\imath$ in cotemporal gauge. Since $d\textsf{A}^\imath$ is static with temporal gauge, $I_{\rm Hopf}$ is reducible. Besides, note that $d\textsf{A}^a d^{-1}\!\textsf{A}^b$ is static. Since $d\textsf{A}^a\textsf{A}^b$ is static with temporal gauge, $d^{-1}(d\textsf{A}^a\textsf{A}^b)$ is also static. So $\Lambda^{ij}$ must be static.
Since $\textsf{B}^\imath$ is in cotemporal gauge, the static $\Lambda^{ij}$ can be taken to be in temporal gauge, so $I_{\rm Hopf}$ is also reducible. Therefore,
\begin{align}
I_{\rm Hopf}\!=\!\frac{1}{(2\pi)^2}\!\!\int_{M}\!\!d\textsf{A}^\imath_R \textsf{B}^\imath_R\,,\,\,\,\,\,\,
I_{\rm BR}\!=\!\frac{1}{(2\pi)^3}\!\!\int_{M}\!\!\Lambda^{ij}_R\textsf{B}^k_R\,,
\end{align}
where $\textsf{A}^\imath_{R}\!\!=\!\!\int_T\textsf{A}^\imath$, $\textsf{B}^\imath_{R}\!\!=\!\!\int_T\textsf{B}^\imath$ and $\Lambda_R^{ij}\!\!=\!\!\Lambda^{ij}|_{t=t_0}$. By using the commutation properties of reduction with exterior derivative,
\begin{align}\label{gamR}
\Lambda^{ij}_R=\sum_{ab}\frac{1}{2}\varepsilon^{ab}\big(2d\textsf{A}^a_Rd^{-1}\!\textsf{A}^b_R+d^{-1} (d\textsf{A}^a_R\textsf{A}^b_R)\big)\,.
\end{align}
Note that $\Sigma^{\imath}$ is static in temporal gauge and $J^\imath$ is in cotemporal gauge, they admit well-defined dimension reduction $\Sigma^{\imath}_R\!\!=\!\!\Sigma^{\imath}|_{t=t_0}\!\!=\!\!\frac{1}{2\pi}d\textsf{A}^{\imath}_R$ and $J^\imath_R\!\!=\!\!\int_{T}\!J^\imath\!\!=\!\!\frac{1}{2\pi}d\textsf{B}_{R}^\imath$. Therefore Eq.\ (\ref{SHOPFS}) and (\ref{SBRS}) follow immediately if we express $I_{\rm Hopf}$ and $I_{\rm BR}$ above in terms of $\Sigma^{\imath}_R$ and $J^\imath_R$.

We now discuss the physical meaning of $\Sigma^{\imath}_R$ and $J^\imath_R$. Given static $\Sigma^\imath\!\!=\!\!\delta(\mathcal{S}^\imath)$ in temporal gauge and $J^\imath\!\!=\!\!\delta(\Gamma^\imath)$ in cotemporal gauge. Let $\omega$ be a static $2$-form in temporal gauge, and $\lambda$ be a $2$-form in cotemporal gauge. Since the loops are static, we have $\SS^\imath\!\!=\!\!m_\imath\!\times\!T$, by using the reduction of spacetime integral,
\begin{align}
\!\int_{M}\!\!\!\lambda\delta(\SS^\imath)|_{t=t_0}
\!=\!\!\int_{\MM}\!\!\!\lambda\delta(\SS^\imath)
\!=\!\!\int_{\SS^\imath}\!\!\!\lambda
\!=\!\!\int_{m_\imath\times T}\!\!\!\!\!\!\lambda
\!=\!\!\int_{M}\!\!\!\lambda\delta(m_\imath).
\end{align}
So $\Sigma^{\imath}_R\!\!=\!\!\delta(\mathcal{S}^\imath)|_{t=t_0}\!\!=\!\!\delta(m_\imath)$ with support on the position of the loop $m_\imath$. Again, by the reduction of spacetime integral,
\begin{align}
\!\int_{M}\!\int_{T}\!\!\!\delta(\Gamma^\imath)\omega
\!=\!\!\int_{\MM}\!\!\!\delta(\Gamma^\imath)\omega
\!=\!\!\int_{\Gamma^\imath}\!\!\!\omega
\!=\!\!\int_{\gamma_{e_\imath}}\!\!\!\omega
\!=\!\!\int_{M}\!\!\!\delta(\gamma_{e_\imath})\omega\,,
\end{align}
where the third equal sign follows from the fact that $\omega$ is static. Therefore, we can write $J^\imath_R\!\!=\!\!\int_{T}\!\delta(\Gamma_{e_\imath})\!\!=\!\!\delta(\gamma_{e_\imath})$ with support on the trajectory $\gamma_{e_\imath}$ of the particle $e_\imath$.


\subsubsection{\,\,\,\,2.2.4 Geometric meaning}\label{intgeo}

Here, we show the geometric meaning of $I_{\rm Hopf}$ and $I_{\rm BR}$ under dimension reduction. Suppose the loops are kept fixed with zero background particle density, and consider the braiding process described by $m_\imath$ and $\gamma_{e_\imath}$ for $\imath\!=\!i,j,k$. Recall that the whole process gauge is invariant iff $m_i$, $m_j$ and $\gamma_{e_k}$ are mutually unlinked circles. We are going to show that the integrals Eq.\ (\ref{SHOPFS}) and (\ref{SBRS}) can be expressed as
\begin{align}
I_{\rm Hopf}=\mathfrak{L}(m_\imath,\gamma_{e_\imath})
\,,\,\,\,\,\,\,I_{\rm BR}=\bar{\mu}(m_i,m_j,\gamma_{e_k})\,.
\end{align}
Hence $I_{\rm Hopf}$ is understood as the Hopf linking number $\mathfrak{L}$, while $I_{\rm BR}$ can is interpreted as the Milnor's triple linking number $\bar{\mu}$ under dimension reduction.

We now discuss the geometric meaning of $I_{\rm Hopf}$ and $I_{\rm BR}$. We first rewrite each of the two integrals in Eq.\ (\ref{SHOPFS}) and (\ref{SBRS}) to a geometrically understandable form by expressing the integrand in term of delta-forms. For $\imath\!=\!i,j,k$, we have
\begin{align}\label{comp1}
\Sigma^\imath_R=\delta(m_{\imath})\,,\,\,\, J^\imath_R=\delta(\gamma_{e_\imath})\,,
\end{align}
which is defined on $M$. Let $\mathsf{S}_{m_\imath}$ and $\mathsf{S}_{e_\imath}$ be the Seifert surfaces bound by $m_\imath$ and $\gamma_{e_\imath}$ respectively. We have
\begin{align}\label{comp2}
d^{-1}\Sigma^\imath_R=\delta(\mathsf{S}_{m_\imath})\,,\,\,\, d^{-1}J^\imath_R=\delta(\mathsf{S}_{e_\imath})\,,
\end{align}
which is also defined on $M$. Now, denote $\gamma^i,\gamma^j,\gamma^k$ as the three mutually unlinked closed curves $m_i$, $m_j$, $\gamma_{e_k}$ respectively binding Seifert surfaces $\mathsf{S}^i$, $\mathsf{S}^j$, $\mathsf{S}^k$. Let $\gamma^\imath_{x}$ be a segment of the closed curve $\gamma^\imath$ from $x_0$ to $x$, where $x_0$ and $x$ are two points on $\gamma^\imath$ for $\imath\!=\!i,j,k$. Along the closed curve $\gamma^\imath$, since $\gamma^{\imath'}$ unlinked with $\gamma^\imath$ for $\imath'\!\neq\! \imath$, we have well-defined $\,d^{-2}\delta(\gamma^{\imath'})\!=\!d^{-1}\delta(\mathsf{S}^{\imath'}\!)\!=\!\int_{\gamma^{\imath}_{x}}\delta(\mathsf{S}^{\imath'}\!)\!=\!\int_{M}\delta(\gamma^{\imath}_{x}\cap\mathsf{S}^{\imath'}\!)\,$ which is a piecewise continuous function in $x$ with unit jump occurring at any of the intersections in $\gamma^{\imath}_{x}\cap\mathsf{S}^{\imath'}$. For $a\!=\!i,j$, we have
\begin{align}\label{comp3}
d^{-2}\Sigma^a_R\!=\!\!\!\int_{M}\!\!\!\delta(\gamma^{\imath}_{x}\cap\mathsf{S}^{a})\,,\,\,\, d^{-2}J^k_R\!=\!\!\!\int_{M}\!\!\!\delta(\gamma^{\imath}_{x}\cap\mathsf{S}^{k})\,,
\end{align}
which is defined on $\gamma^{\imath}$, where $\imath\neq a$ for the former and $\imath\neq k$ for the later. Now we have all the ingredients for the geometric interpretation. By using Eq.\ (\ref{comp1}) and (\ref{comp2}),   Eq. (\ref{SHOPFS}) becomes
\begin{align}
I_{\rm Hopf}=
\int_{M}\!\!\!\delta(\mathsf{S}_{m_\imath}\!\!\cap\!\gamma_{e_\imath})\,.
\end{align}
Since $\int_{M}\!\delta(\mathsf{S}_{m_\imath}\!\!\cap\!\gamma_{e_\imath})$ is the sum of signed intersections of $\mathsf{S}_{m_\imath}$ and $\gamma_{e_\imath}$,
$I_{\rm Hopf}$ can be interpreted as the Hopf linking number $\mathfrak{L}$ between the loop $m_\imath$ and  the particle trajectory $\gamma_{e_\imath}$.
Likewise, by using Eq.\ (\ref{comp1}), (\ref{comp2}) and (\ref{comp3}),   Eq.~(\ref{SBRS}) becomes
\begin{align}
I_{\rm BR}\!=&\!\!\!
\int_{M}\!\!\!\delta(\mathsf{S}^i\!\cap\!\mathsf{S}^j\!\cap\!\mathsf{S}^k)\nonumber\\
&-
\sum_{\imath_{\textrm{\tiny1}}\imath_{\textrm{\tiny2}}\imath_{\textrm{\tiny3}}}\frac{1}{2}
\varepsilon^{\imath_{\textrm{\tiny1}}\imath_{\textrm{\tiny2}}\imath_{\textrm{\tiny3}}}\!\!\!
\int_{M}\!\!\!\delta(\gamma^{\imath_\textrm{\tiny3}}\!\!\cap\!\mathsf{S}^{\imath_\textrm{\tiny1}})\!\!
\int_{M}\!\!\!\delta(\gamma^{\imath_\textrm{\tiny3}}_{x}\!\!\cap\!\mathsf{S}^{\imath_\textrm{\tiny2}})\,,\nonumber
\end{align}
where $\imath_1,\imath_2,\imath_3=i,j,k$. Notice that the term $\int_{M}\delta(\mathsf{S}^i\cap \mathsf{S}^j\cap\mathsf{S}^k)$ is the sum of the signed intersections of $\mathsf{S}^i$, $\mathsf{S}^j$ and $\mathsf{S}^k$, that is, $t_{ijk}$. Also, observe that the integral $\int_{M}\delta(\gamma^{\imath_3}\cap\mathsf{S}^{\imath_1})\!\int_{M}\delta(\gamma^{\imath_3}_{x}\cap\mathsf{S}^{\imath_2})$
is the sum of signed occurrences of $\mathsf{S}^{\imath_1}$ after $\mathsf{S}^{\imath_2}$ along $\gamma^{\imath_3}$, that is, $e_{\imath_1\imath_2\imath_3}$. Since $e_{\imath_1\imath_2\imath_3}$ is anti-symmetric \cite{mellor2003geometric}, the integral $I_{\rm BR}$ can be geometrically interpreted as $t_{ijk}\!-\!(e_{ijk}\!+\!e_{jki}\!+\!e_{kij}\!)$ which is precisely the Milnor's triple linking number \cite{milnor1954link} of the closed curves $\gamma^i$, $\gamma^j$, $\gamma^k$, or equivalently, $m_i$, $m_j$, $\gamma_{e_k}$.

\section{Part 3: Implied SPT phases}\label{appendix_mixture_SPT}

In this part, we discuss briefly the SPT phases implied by the BR braiding statistics. Basically, the BR braiding statistics implies a class of highly unexplored SPT phases protected by \emph{mixed} global symmetry $G\!=\!\prod_i\Z_{N_i}$ where some components are the usual symmetries acting on particles, whereas the others are symmetries acting on loops. We call these phases mixed SPT phases. In section 3.1, we introduce the mixed SPT phases. In section 3.2, we present effective field theories for the mixed SPT phases. In section 3.3, we give a condensation picture for these exotic SPT phases.

\subsection{3.1 Mixed SPT Phases}

SPT phases are short-range entangled phases of matter protected by a global symmetry \cite{Chenlong,PhysRevB.85.075125,1DSPT,Chen_science}. Exhibiting trivial braiding statistics in the bulk, they manifest non-trivial electromagnetic response under external fields. There is a duality correspondence between SPT phases and topologically ordered phases \cite{levin_gu_12,wang_levin1}. Promoting the external fields in an SPT phase to dynamical gauge fields leads to a discrete gauge theory with certain braiding data. Conversely, suppressing the gauge fluctuations recovers the SPT phase. In particular, freezing the cyclic gauge fields $A^i, A^j$ and $B^k$ in TQFTs with BR braiding statistics leads to SPT phases with $\Z_{N_i}\!\times\!\Z_{N_j}\!\times\!\Z_{N_k}$ global symmetry. In contrast to the traditional SPT phases in which the global symmetry acts on either point-like or loop-like charges \cite{Chenlong,Chen_science,string7}, the 1-forms $A^i$, $A^j$ couple to point-like $\Z_{N_i},\Z_{N_i}$ charges respectively whereas the 2-form $B^k$ couple to loop-like $\Z_{N_k}$ charges. We call these as \emph{mixed SPT phases}, in which microscopic degrees of freedom consist of both particles and loops.
These mixed SPT phases are characterized by their electromagnetic response
\begin{align}\label{response}
S_{\rm Resp}=S_{\rm AAB}=\int\!\frac{n k_{ij,k} }{(2\pi)^2N_{ijk}} A^i A^j B^k\,
\end{align}
under the closed cyclic external fields $A^i\!, A^j$ and $B^k$. It means that the intersection of two domain walls binding $\frac{2\pi}{N_i}$ flux in $A^i$ and a $\frac{2\pi}{N_j}\!$ flux in $A^j\!$ traps \small$\frac{k_{ij,k}N_k}{N_{ijk}}$\normalsize amount of line charges of $B^k$, which is defined mod $N_k$. Therefore, the mixed SPT phases are also classified by $k_{ij,k}\!\in\!\Z_{N_{ijk}}$, which characterizes the BR braiding statistics in the dualized picture.

\subsection{3.2 Effective field theoretical description}

Here, we introduce the effective field theories \cite{YeGu2015} for the mixed SPT phases with global symmetry $\Z_{N_i}\!\times\!\Z_{N_j}\times\!\Z_{N_k}$, where $\Z_{N_a}\!$ acts on particles for $a\!=\!i,j$ and $\Z_{N_k}\!$ acts on loops. The response action is $S_{\rm AAB}\!$ in which $A^a\!$ and $B^k\!$ are closed cyclic probe fields. Since $S_{\rm AAB}$ is large gauge invariant, following the functional bosonization scheme \cite{PhysRevB.87.085132, PhysRevB.93.155122}, we have
\begin{align}\label{spt_field}
\!\!\!\!S_{\rm SPT}=&\int \sum_{\imath}\frac{1}{2\pi}b^\imath da^\imath+\frac{n c_{ij,k}}{(2\pi)^3}a^i a^j b^k\nonumber\\
&\,-\sum_{a}\frac{1}{2\pi}A^adb^a-\frac{1}{2\pi}B^kda^k~,
\end{align}
where $a^i\!, a^j\!, a^k\!$ and $b^i\!, b^j\!, b^k\!$ are some dynamical compact U(1) $1$-form and $2$-form gauge fields respectively. In particular, $\frac{1}{2\pi}db^a$ describes the point charge of $A^a$ and $\frac{1}{2\pi}da^k$ describes the line charge of $B^k$ \cite{string7,kap_13}. The action $S_{\rm SPT}$, which manifests gauge invariant in a way similar to the TQFTs in Eq.\ \!(\ref{ATQFT}), exhibits trivial braiding statistics in the bulk and correctly reproduces the electromagnetic response action $S_{\rm AAB}$ for the mixed SPT phases. With the effective field theories, the duality correspondence \cite{levin_gu_12,YW12,corbodism3,YeGu2015,bti1,Ye14b,PhysRevB.86.125119,PhysRevB.93.115136,PhysRevLett.112.141602,ye16_set,YeGu2015,2016arXiv161008645Y,PhysRevB.87.085132,PhysRevB.93.155122,ryu_zhang_2012,PhysRevB.88.075125,2017arXiv170508911C,Hung_Wen_gauge} can be made precise.
Starting with the hydrodynamic description $S_{\rm SPT}\!$ of the mixed SPT phases, we get the response $S_{\rm AAB}\!$ by integrating out the dynamical gauge fields. By promoting $A^a$ and $B^k$ to dynamical gauge fields, together with the incorporation of $S_{\rm BF}$ for the local flatness conditions and cyclic constraints of $A^a\!$ and $B^k\!$, we get the TQFTs in Eq.\ (\ref{ATQFT}) with BR braiding statistics.

\subsection{3.3 Condensation picture}

Here, we develop a condensation picture for the mixed SPT phases. SPT phases can be understood as a result of decorated domain wall proliferation in symmetry-breaking condensates \cite{PhysRevB.93.115136,YeGu2015,Chen:2014aa}. Here, by proliferating defects in a condensate with two flavors of bosonic particles and one flavor of loops, we get at SPT phases with global symmetry $\Z_{N_i}\!\times\!\Z_{N_j}\times\!\Z_{N_k}$, where $\Z_{N_a}\!$ acts on particles for $a\!=\!i,j$ and $\Z_{N_k}\!$ acts on loops.

Consider a mixture of two species of boson condensates and one species of loop condensate described by the $U^{3}(1)$ non-linear $\sigma$-model with a multi-kink term,
\begin{align}
S_{\sigma}\!=\!\int\!\sum_{a}\frac{g_a}{2}d\theta^a\!\!\star\!d\theta^a
\!+\!\frac{g_k}{2}d\Theta^k\!\!\star\!d\Theta^k\!+\!\frac{n c_{ij,k}}{(2\pi)^3}d\theta^i d\theta^j d\Theta^k\,,\nonumber
\end{align}
where the $0$-form $\theta^a$ and $1$-form $\Theta^k$ describe the phase fluctuation for the particle and the loop condensate respectively. In the disordered regime where the domain walls condense, the system exhibits the discrete global symmetry $\Z_{N_i}\!\times\!\Z_{N_j}\!\times\! \Z_{N_k}$ if $c_{ij,k}\!=\!\frac{2\pi k_{ij,k}}{N_{ijk}}\!$. In such case, the phase fluctuations admit a smooth and a singular part, $d\theta^a\!=\!d\theta^a_s\!+\!a^a$ and $d\Theta^k\!=\!d\Theta^k_s\!+\!b^k\!$, where the 1-forms $a^a$ and 2-form $b^k$ are compact.
Under the standard duality, $\theta^a_s$ and $\Theta^k_s$ are integrated out, and we get
\begin{align}
S_{\rm SPT}=\int \sum_{\imath}\frac{1}{2\pi}b^\imath da^\imath+\frac{n c_{ij,k}}{(2\pi)^3}a^i a^j b^k~,
\end{align}
where $b^a$ and $a^k$ are  respectively compact 2-forms and 1-form gauge fields which appear under the duality procedure. Also, $\frac{1}{2\pi} db^a$ is the point charge of the ${\Z}_{N_a}$ symmetry and $\frac{1}{2\pi} da^k$ is the line charge of the ${\Z}_{N_k}$ symmetry. Adding back the coupling term with the probes $A^a$ and $B^k$, we get precisely the theories describing the mixed SPT phases in Eq.\ (\ref{spt_field}). As a side, it will be interesting to construct SPT phases when there are many coexisting topological terms such as $aada$ and $aaaa$ \cite{YeGu2015,corbodism3,ye16_set,2016arXiv161008645Y,string4,PhysRevLett.114.031601,3loop_ryu,string10,2016arXiv161209298P,Tiwari:2016aa,PhysRevB.95.035131},  $bb$ \cite{horowitz89,bti2,Kapustin2014,string7,Keyserlingk13,Walker2012},  $dada$  \cite{PhysRevX.6.011034,ye16a,YW13a,Ye:2017aa,bti6,lapa17} on top of the $aab$ topological term.

\end{document}